%% file: main_arxiv.tex
\documentclass[acmtog]{acmart}

\def\ie{\emph{i.e.}}
\def\eg{\emph{e.g.}}
\def\wrt{w.r.t.}

\usepackage{booktabs} %

\usepackage{multirow}
\usepackage{xspace}
\usepackage{bm}
\usepackage{pifont}
\usepackage{cleveref}
\usepackage{subcaption}

\newcommand{\dataset}{Anymate\xspace}
\newcommand{\datasetacronym}{Anymate\xspace}

\copyrightyear{2025}
\acmYear{2025}
\acmConference[SIGGRAPH Conference Papers '25]{Special Interest Group on Computer Graphics and Interactive Techniques Conference Conference Papers }{August 10--14, 2025}{Vancouver, BC, Canada}
\acmDOI{10.1145/3721238.3730743}
\acmISBN{979-8-4007-1540-2/2025/08}

\begin{document}

\title{Anymate: A Dataset and Baselines for Learning 3D Object Rigging}

\author{Yufan Deng}
\authornote{Equal contribution. The order of authorship was determined alphabetically. Work was done when Y. Deng and Y. Zhang were visiting students at Stanford University; they are currently with the Hong Kong University of Science and Technology.}
\affiliation{%
\institution{Stanford University}
 \city{Stanford}
 \country{USA}
}

\author{Yuhao Zhang}
\authornotemark[1]
\affiliation{%
\institution{Stanford University}
 \city{Stanford}
 \country{USA}
}

\author{Chen Geng}
\affiliation{%
 \institution{Stanford University}
 \city{Stanford}
 \country{USA}
}

\author{Shangzhe Wu}
\authornote{Equal advising.}
\affiliation{%
 \institution{Stanford University}
 \city{Stanford}
 \country{USA}
}
\affiliation{%
 \institution{University of Cambridge}
 \city{Cambridge}
 \country{UK}
}

\author{Jiajun Wu}
\authornotemark[2]
\affiliation{%
 \institution{Stanford University}
 \city{Stanford}
 \country{USA}
}

\begin{teaserfigure}
\begin{center}
\includegraphics[width=\linewidth]{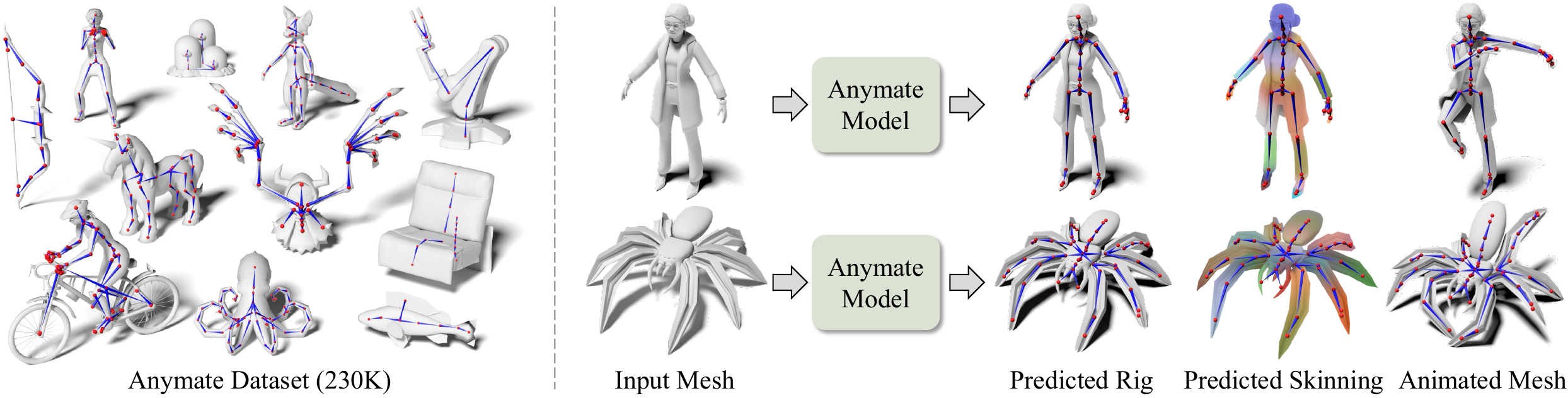}
\end{center}
\vspace{-1em}
\captionsetup{type=figure}
\captionof{figure}{
We present the \textbf{\dataset Dataset} and a learning-based framework for automatic 3D object rigging.
This dataset comprises 230K 3D assets with expert-crafted rigging and skinning information.
Using this dataset, our framework learns to predict bone skeletons and skinning weights automatically from a given 3D mesh, allowing users to create realistic animations by manipulating the predicted skeleton.
}%
\label{fig:teaser}
\end{teaserfigure}

\begin{abstract}
Rigging and skinning are essential steps to create realistic 3D animations, often requiring significant expertise and manual effort. Traditional attempts at automating these processes rely heavily on geometric heuristics and often struggle with objects of complex geometry. Recent data-driven approaches show potential for better generality, but are often constrained by limited training data. We present the \textbf{\dataset Dataset}, a large-scale dataset of 230K 3D assets paired with expert-crafted rigging and skinning information---70 times larger than existing datasets. Using this dataset, we propose a learning-based auto-rigging framework with three sequential modules for joint, connectivity, and skinning weight prediction. We systematically design and experiment with various architectures as baselines for each module and conduct comprehensive evaluations on our dataset to compare their performance. Our models significantly outperform existing methods, providing a foundation for comparing future methods in automated rigging and skinning. Code and dataset can be found at \url{https://anymate3d.github.io/}.
\end{abstract}
\vspace{-1.5em}



\maketitle

\section{Introduction}

With rapid advances of immersive\footnotetext[1]{Equal contribution, alphabetical order. $^\dagger$Equal advising.}
 media comes a growing demand for automated 3D content creation.
3D animation is essential nowadays across the entertainment industry, from game design and movie production to a wide range of AR/VR applications.
Yet, producing realistic 3D animations remains one of the most labor-intensive tasks, often requiring enormous expert effort.

Creating an animation of a 3D object generally involves two steps.
The first is to describe the desired motion using a set of sparse, interpretable motion handles, such as keypoints or bones---often referred to as a ``rig.''
The second step, known as ``skinning,'' then defines how the 3D object deforms densely according to the movement of these handles.
Both steps require specialized expertise and meticulous effort to ensure realistic animation results.

In this paper, our goal is to develop a model that can efficiently and fully automatically turn a static 3D asset into an animatable version.
To this end, we propose a learning-based auto-rigging system that generates a rig and the corresponding skinning mechanism from a given 3D mesh.
This allows users to create realistic 3D animations by simply manipulating the underlying rig in an intuitive manner.

The key insight is to leverage existing large-scale animated 3D assets handcrafted by expert artists.
Large-scale 3D data has proven crucial in training high-quality 3D generation models \cite{shi2023zero123++,liu2024one}.
Notably, the Objaverse datasets \cite{objaverse,deitke2024objaverseXL}, aggregating over 10 million 3D assets from various public sources, have powered many recent state-of-the-art methods.
However, prior efforts primarily use such assets for static 3D generation, neglecting the fact that many of these assets also come with valuable animation information skillfully crafted by the artists.

To train a learning-based model for 3D object rigging, we assemble a large-scale dataset of 
230K rigged 3D assets curated from the Objaverse-XL Dataset~\cite{deitke2024objaverseXL}, each paired with artist-created rigging and skinning in a unified format, dubbed the \emph{\dataset Dataset}.
This dataset is 70 times larger than existing public rigging datasets, as summarized in Table~\ref{tab:dataset}, and
contains a wide spectrum of 3D objects, ranging from humanoid and animal characters to animated everyday objects like furniture and machines.

Using the \dataset Dataset, we develop a learning-based framework that predicts a 3D bone skeleton and corresponding skinning weights fully automatically from any given 3D object mesh.
This framework consists of three sequential modules for joint, connectivity, and skinning weight prediction.
For each module, we design several architectures as strong baselines and provide a comprehensive set of evaluations and comparisons on the proposed dataset.
These experiments show that the larger scale of training data significantly improves prediction results.
Moreover, our proposed architectures demonstrate better scalability, outperforming existing methods by a considerable margin.
The dataset, code, and pretrained weights will be publicly released to facilitate future research.

Our contributions are summarized as follows:
\begin{enumerate}
    \item We introduce the \dataset Dataset consisting of 230K 3D assets with rigging and skinning information.
    \item We develop a learning-based framework for automatic 3D object rigging and skinning.
    \item We design a comprehensive set of baselines with a variety of architectures and provide thorough evaluations and comparisons on the proposed dataset, providing a reference point for future comparisons.
\end{enumerate}

\section{Related Work}

\subsection{3D Object Rigging Datasets}
Auto-rigging has traditionally been approached through geometry-based optimization.
Hence, limited efforts were dedicated to collecting and open-sourcing large-scale rigging datasets before the recent rise of data-driven methods.
Most of the existing public rigging datasets focus primarily on humanoid and animal characters, such as SMPL~\cite{loper2023smpl}, DeformingThings4D~\cite{li20214dcomplete}, Maximo~\cite{mixamo}, Planet Zoo~\cite {wu2022casa}, Model Resource~\cite{xu2019predicting}, and RaBit \cite{luo2023rabit}.
The largest of these, Model Resource, contains only 3.3K assets.
Our dataset consists of 230K assets---around 70x larger---which is crucial for training robust learning-based models.
Table~\ref{tab:dataset} compares the key features of these datasets.

Our \dataset Dataset is built upon the recently released Objaverse-XL Dataset~\cite{deitke2024objaverseXL}, which consists of over 10M 3D assets.
The authors of \cite{li2024puppetmaster} also extract a set of 14K animated assets in Objaverse~\cite{objaverse}, dubbed Objaverse-Animation.
Unlike ours, this dataset is curated to train an implicit image-based model for object animation, offering no direct access to rigging and skinning information.

\begin{table}[t]
    \small
    \centering
    \caption{\textbf{3D Object Rigging Datasets.} Existing datasets are limited in both size and variety. In contrast, our \dataset Dataset is 70 times larger than the existing datasets with rigging. 
    }
        \resizebox{\linewidth}{!}{%
    \begin{tabular}{lccc}
        \toprule
        Dataset & Size & Category & Rigging \\  
        \midrule
        DeformingThings4D \cite{li20214dcomplete}      & 2.0K & humanoid, animal & \textcolor{red}{\ding{55}} \\
        Objaverse-Animation \cite{li2024puppetmaster}      & 14K & generic & \textcolor{red}{\ding{55}} \\
        SMPL \cite{loper2023smpl}                    & 0.2K & human &\textcolor{green}{\ding{51}} \\
        Mixamo \cite{mixamo}                    & 0.1K &humanoid &\textcolor{green}{\ding{51}}\\
        Planet Zoo \cite{wu2022casa}            & 0.25K &animal & \textcolor{green}{\ding{51}}\\
        RaBit \cite{luo2023rabit}               & 1.5K & humanoid & \textcolor{green}{\ding{51}}\\

        Model Resource \cite{xu2020rignet}      & 3.3K &humanoid, animal & \textcolor{green}{\ding{51}} \\
        \midrule
        \dataset Dataset (ours)                      & 230K & generic & \textcolor{green}{\ding{51}}\\
        \bottomrule
    \end{tabular}
    }

    \label{tab:dataset}
\end{table}

\subsection{Skeleton-based Automatic 3D Object Rigging}
\label{rela:rig}

Early auto-rigging methods, such as Pinocchio \cite{baran2007automatic} and Avatar Reshaping \cite{feng2015avatar}, fit a template skeleton to a given mesh through optimization.
This requirement of a template skeleton heavily limits the range of applicable objects, typically to humanoid or quadruped characters.
Several works have explored data-driven approaches for rigging prediction~\cite{xu2020rignet,xu2019predicting,ma2023tarig}.

In particular, \cite{xu2019predicting} uses a 3D convolution-based architecture that operates on voxelized shapes, whereas the follow-up works of RigNet~\cite{xu2020rignet} and Tarig~\cite{ma2023tarig} replace it with graph neural networks that directly process mesh vertices.
However, these models are all trained on small-scale datasets with less than 3K objects and often generalize poorly to complex objects, as shown in \cref{fig:result}. 

Concurrent to our work, RigAnything~\cite{liu2025riganything} and MagicArticulate~\cite{song2025magicarticulate} present auto-regressive rigging models learned from large-scale data. Compared to those works, we focus on presenting the dataset, benchmark, and a set of baseline methods for future comparison.

\subsection{Non-skeletal Automatic 3D Object Rigging}

There also exist works that achieve automatic rigging using representations other than skeleton. We summarize them as handle-based, cage-based, and neural-based methods.
\paragraph{Handle-based.} KeypointDeformer~\cite{jakab2021keypointdeformer} and DeepMetaHandles~\cite{liu2021deepmetahandles} pioneer the unsupervised learning of deformation handles. While being annotation-free, these methods often require datasets with minor deformations or aligned poses, which limits their applications to broader categories. In contrast, skeleton-based approaches benefit from large-scale artist-annotated ground truth for supervised learning. We have tried KeypointDeformer~\cite{jakab2021keypointdeformer} but find that instead of manipulating motion, handle-based methods focus on manipulating the mesh's shape, i.e, adjusting an airplane's wing length. 
\paragraph{Cage-based.} Methods like Neural Cages~\cite{yifan2020neural} offer promising approaches to learning rigging without large annotated datasets. However, they struggle with objects with complex hierarchical kinematic structures and large-scale deformations.
\paragraph{Neural-based.}  Rigging methods with a neural representation~\cite{aigerman2022neural,qin2023neural} are promising for modeling complex motions like faces, but often require an extensive amount of observations under known poses for training. Moreover, it remains challenging to obtain an interpretable neural representation. 

In summary, different rigging modalities serve distinct purposes. The skeleton-based approach benefits from large-scale annotated datasets, applies to diverse categories, and integrates seamlessly with traditional graphics pipelines. Other modalities excel in specific contexts, such as with aligned datasets or synthetic data.

\subsection{Automatic Skinning}
\label{rela:skin}

In 3D animation, artists usually need to paint skinning weights on a mesh to indicate how the vertices should deform in response to bone movement.
Techniques like Linear Blend Skinning (LBS)~\cite{magnenat1988joint} are then used to drive the mesh based on bone animations.
Various geometric methods have been proposed to automatically determine skinning weights~\cite{bang2018spline, dionne2014geodesic, dionne2013geodesic, chen2021snarf, kavan2012elasticity, jacobson2011bounded, song2024automatic}, such as using Laplacian energy \cite{jacobson2011bounded} or geodesic distance \cite{dionne2013geodesic}.
However, these methods often struggle with objects of complex geometry.

Previous data-driven approaches have shown promise for better generality~\cite{xu2020rignet,mosella2022skinningnet,ma2023tarig,liu2019neuroskinning}.
Nevertheless, similar to rigging, previous methods either focus on specific categories and rely on templates~\cite{liao2024vinecs} or train on limited datasets~\cite{xu2020rignet, mosella2022skinningnet}.

\section{\dataset Dataset}

To learn 3D object auto-rigging, a sizable collection of examples is essential.
We introduce the \emph{\dataset Dataset}, a large-scale dataset of manually crafted animated 3D objects, building upon the existing Objaverse-XL Dataset \cite{deitke2024objaverseXL}.
In the following, we first give an overview of the representation for 3D object animation, and then describe the procedures used to build the \dataset Dataset.

\subsection{Preliminary: 3D Object Animation}
\label{data:define}

When artists create an animation of a 3D object, two key components are typically required: rigging and skinning.
The former establishes a set of sparse, interpretable handles that serve as a scaffold to outline the intended motion.
The latter, in turn, defines how the 3D object should deform densely in response to the motion scaffold.
In modern computer graphics, the rig is often represented by a skeleton of bones, and the skinning is usually defined by weights associating mesh vertices to these underlying bones.

Concretely, given a 3D mesh with vertices denoted by $\mathbf{V} \subset \mathbb{R}^3$, we first define a set of bones, indexed by $b = 1, ..., B$.
Each bone is an oriented line segment specified by a pair of head and tail joints $\mathbf{J}_b^\text{head}, \mathbf{J}_b^\text{tail} \in \mathbb{R}^3$ in the world coordinate space.
In practice, these bones are typically connected via shared joints, forming a kinematic tree structure.

To animate the object, we then deform each vertex $\mathbf{V}_i$ of the mesh according to the poses $\xi$ of these bones via the Linear Blend Skinning (LBS) equation \cite{magnenat1988joint,Lewis2000skinning}:
\begin{equation}
\begin{aligned}
\label{eq:skinning}
    \mathbf{V}_i'(\xi) =\left( \sum_{b=1}^B w_{i,b} G_b(\xi) G_b(\xi^*)^{-1} \right) \mathbf{V}_i, \\
    G_1 = g_1,
    ~~
    G_b = G_{\pi(b)} \circ g_b,
    ~~
    g_b(\xi_b) \in SE(3).
\end{aligned}
\end{equation}
Here, $\xi = \{\xi_1, ..., \xi_B\}$ denotes the poses of the bones in their local coordinate frames, and $\xi^*$ denotes the rest pose, which can be derived from the joint locations $\mathbf{J}_b^\text{head}$ and $\mathbf{J}_b^\text{tail}$ (see the supp.~mat.). $G$
 stands for the global transformation matrix after composing all the local transformations $g_b$
 along the kinematic chain.
$\pi(b)$ retrieves the index of the parent of bone $b$ in the kinematic tree.
$w_{i,b}$ is the \emph{skinning weight} of vertex $\mathbf{V}_i$ with respect to bone $b$.
Essentially, for each bone $b$, vertex $\mathbf{V}_i$ in the world coordinate frame is first transformed to the local coordinate frame of the bone following the inverse kinematic chain $G_b^{-1}$, and then reposed together with the bone following the forward kinematic chain $G_b$.
The final position of the vertex $\mathbf{V}_i'$ is the sum of transformations from all bones weighted by $w_{i,b}$.

Therefore, to animate a mesh, we need a set of bones $\mathcal{B}$, each specified by a pair of joints $\mathcal{B}_b = (\mathbf{J}_b^\text{head}, \mathbf{J}_b^\text{tail})$, and a function $\mathcal{S}$ that returns the skinning weights for a given point on the surface.
The goal of our model is thus to predict these bones and skinning weights from a given mesh.

\begin{figure}[t]
    \centering
    \begin{subfigure}{0.49\linewidth}
        \centering
        \includegraphics[trim={0pt 10pt 0pt 0pt}, clip, width=\linewidth]{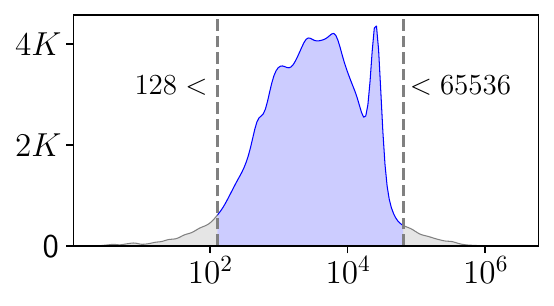}
        \caption{Number of Vertices}
        \label{subfig:vertex}
    \end{subfigure}
    \hfill
    \begin{subfigure}{0.5\linewidth}
        \centering
        \includegraphics[trim={0pt 10pt 0pt 0pt}, clip, width=\linewidth]{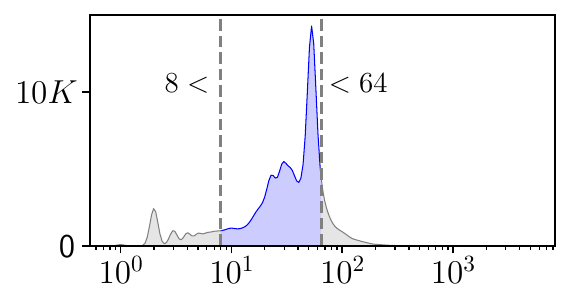}
        \caption{Number of Bones}
        \label{subfig:bone}
    \end{subfigure}
    \begin{subfigure}{\linewidth}
        \centering
        \vspace{0.5em}
        \includegraphics[width=\linewidth]{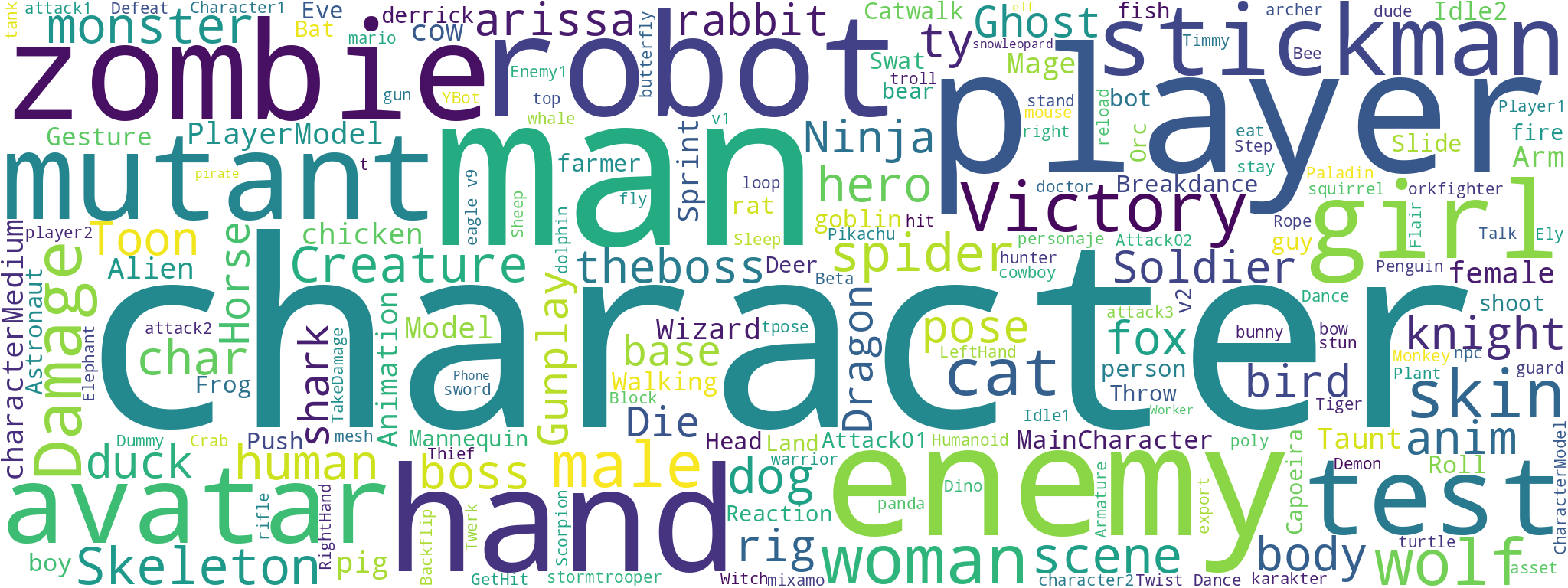}
        \caption{Objects in the \dataset Dataset}
        \label{subfig:word}
    \end{subfigure}
    \vspace{-2em}
    \caption{\textbf{Statistics of the \dataset Dataset.} We filter out assets with a vertex count beyond $(128, 65535)$ and a bone count beyond $(8, 64)$, as shown in (a) and (b).
    The remaining assets span a wide spectrum of objects, as visualized in the word cloud in (c).}
    \label{fig:dataset}
    \vspace{-1.5em}
\end{figure}

\begin{figure*}
    \centering
    \includegraphics[clip, width=\linewidth]{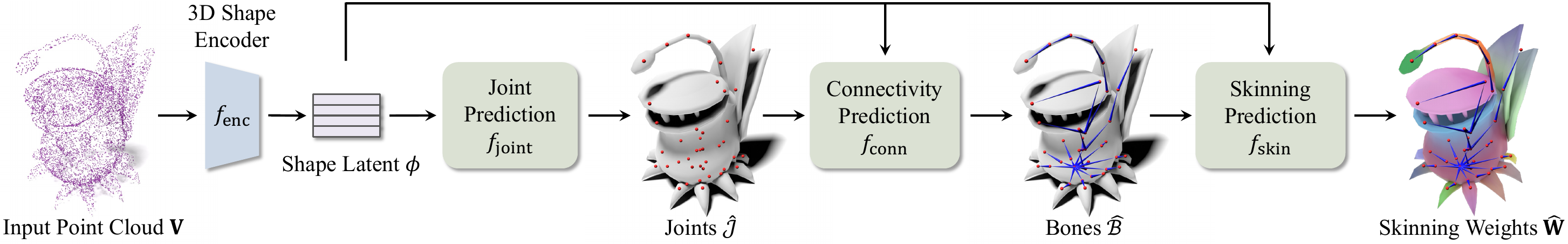}
    \vspace{-1.8em}
    \caption{\textbf{Overview of Our Learning-based Auto-rigging Framework.}
    Given an input 3D mesh, we first sample surface points $\mathbf{V}$ and extract shape latent features $\phi$.
    Three sequential modules then predict a set of joints $\hat{\mathcal{J}}$, joint connectivity to establish bones $\hat{\mathcal{B}}$, and finally the skinning weights with respect to each bone $\hat{\mathbf{W}}$, all based on the shape latent.
    See \cref{sec:exp} for details.
    }
    \label{fig:overview}
    \vspace{-1em}
\end{figure*}

\subsection{Dataset Processing}
\label{data:process}

To train such a model, we prepare a large dataset of rigged 3D objects.
Our dataset contains 230K 3D assets of common objects from the Objaverse-XL dataset~\cite{deitke2024objaverseXL}.
Each asset comes with a 3D mesh, a bone skeleton, and skinning weights for each mesh vertex in a unified format.

\paragraph{Filtering.}
Objaverse-XL provides links to over 10 million 3D assets from various public sources in various formats.
We identify the assets that potentially contain rigging and skinning information with a file format of \textsl{.fbx}, \textsl{.dae}, \textsl{.gltf}, or \textsl{.glb}.
While Objaverse-XL reports 438K objects with armatures, we find only 256K that are still publicly available.
We then filter the assets, retaining roughly 200K with a vertex count of $128<|\mathbf{V}|<65,535$ and a bone count of $8<|\mathcal{B}|<64$. Fig.~\ref{fig:dataset} gives an illustration of the statistics. After this, we eliminate redundant data based on skeleton similarity. Next, we remove assets where most of the skeleton ($>70\%$) lies outside the mesh. 
Many assets contain animation information, which is used to augment the dataset. We use no more than three extra key-frame poses from the same asset.
After augmentation, we unify the data format and fix bones and joints. More details can be found in the supp.~mat.

\paragraph{Pre-processing.}
\label{subset:preprocessing}
We load these assets into Blender\footnote{\small\url{https://www.blender.org/}} and extract the mesh, armature (bones), and skinning weights.
For training efficiency, we resample 8,192 points on the mesh, and recompute their skinning weights using barycentric interpolation based on the original annotated skinning weights at the vertices.
Finally, for each asset, we export the original mesh in a unified \textsl{.obj} format, as well as the resampled point clouds and the corresponding bone locations and skinning weights.

\paragraph{Dataset Statistics.}
\label{data:stat}
The final dataset contains $230,716$ assets.
We randomly select 5.6K instances as the test set (\datasetacronym-test), and use the rest 225K for training (\datasetacronym-train).
To facilitate efficient ablation experiments, we also construct a smaller 14K training set (\datasetacronym-small) with a subset of training instances that are included in the original Objaverse 1.0 Dataset~\cite{objaverse}, which is likely of higher quality.

\section{Learning to Rig 3D Objects}
\label{sec:exp}

Using the \dataset Dataset, we train learning-based models to predict bone skeleton and skinning weights from a given 3D object mesh.
These allow users to create 3D animations of the object by simply manipulating the skeleton.

We divide this task into three sequential sub-tasks: joint prediction, connectivity prediction, and skinning weight prediction, as outlined in Fig.~\ref{fig:overview}.
Starting with an input 3D mesh, we first use an encoder $f_\text{enc}$ to extract shape features, which are then fed into the three subsequent prediction modules.
The first module $f_\text{joint}$ predicts a set 3D joints.
The second module $f_\text{conn}$ then connects these joints into bones by predicting binary connectivity for each pair of joints, forming a skeleton.
Finally, the skinning module $f_\text{skin}$ estimates the skinning weight between each mesh vertex and each bone, based on the constructed skeleton.
For each sub-task, we experiment with various architectures as baselines and conduct evaluations on our \dataset Dataset, comparing against previous methods.

In the following, we first introduce the shape encoder (\cref{subsec:encoder}) shared across all prediction modules, and then discuss each individual module (\cref{subsec:joint,subsec:connectivity,subsec:skin}), followed by visual results generated using the predictions (\cref{subsec:visual}).

\subsection{Shape Encoding}
\label{subsec:encoder}

Given a 3D mesh, we first uniformly resample $8,192$ points across its surface and extract shape latent features $\phi$ using an encoder $f_\text{enc}$.
We experiment with two popular transformer-based point cloud encoders, \textbf{Michelangelo}~\cite{zhao2024michelangelo} and \textbf{Point-BERT}~\cite{yu2022point}. 
Michelangelo trains a perceiver-based transformer with learnable query tokens and several self-attention blocks.
Point-BERT trains a discrete Variational Auto-Encoder (dVAE) which first maps a point cloud into a set of patch tokens and subsequently mixes them via a transformer.
Both models encode the point cloud into a set of feature tokens (257 and 513 respectively).
In our experiments, we load the pre-trained weights from existing models and fine-tune them with each subsequent prediction module individually.
Specifically, we load the Michelangelo model pre-trained on the ShapeNet Dataset~(50K)~\cite{chang2015shapenet}, and the Point-BERT model pre-trained on the Cap3D Dataset~(660K)~\cite{luo2024scalable} released by PointLLM~\cite{xu2024pointllm}, as Cap3D is also derived from Objaverse~\cite{objaverse}.

\subsection{Joint Prediction}
\label{subsec:joint}

After obtaining the shape latent $\phi$, the joint prediction module $f_\text{joint}$ predicts a set of joints $\hat{\mathcal{J}} = f_\text{joint}(\phi) = \{\hat{\mathbf{J}}_k\}_{k=1}^K$, where each joint $\hat{\mathbf{J}}_k \in \mathbb{R}^3$ is a 3D coordinate in world space, and the number of joints $K$ may vary across instances.

\begin{figure}[t]
    \centering
    \begin{subfigure}[b]{\linewidth}
        \centering
        \includegraphics[width=0.85\linewidth]{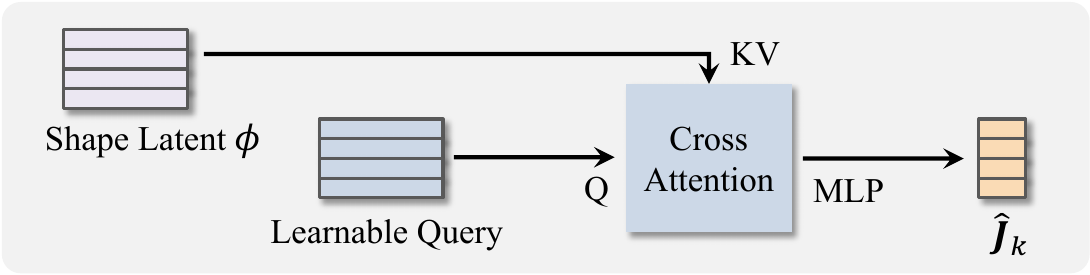}
        \caption{Regression-based Architecture}
        \label{subfig:regress}
    \end{subfigure}
    \begin{subfigure}[b]{\linewidth}
        \centering
        \includegraphics[trim={0pt 0pt 10pt 0pt}, clip, width=0.85\linewidth]{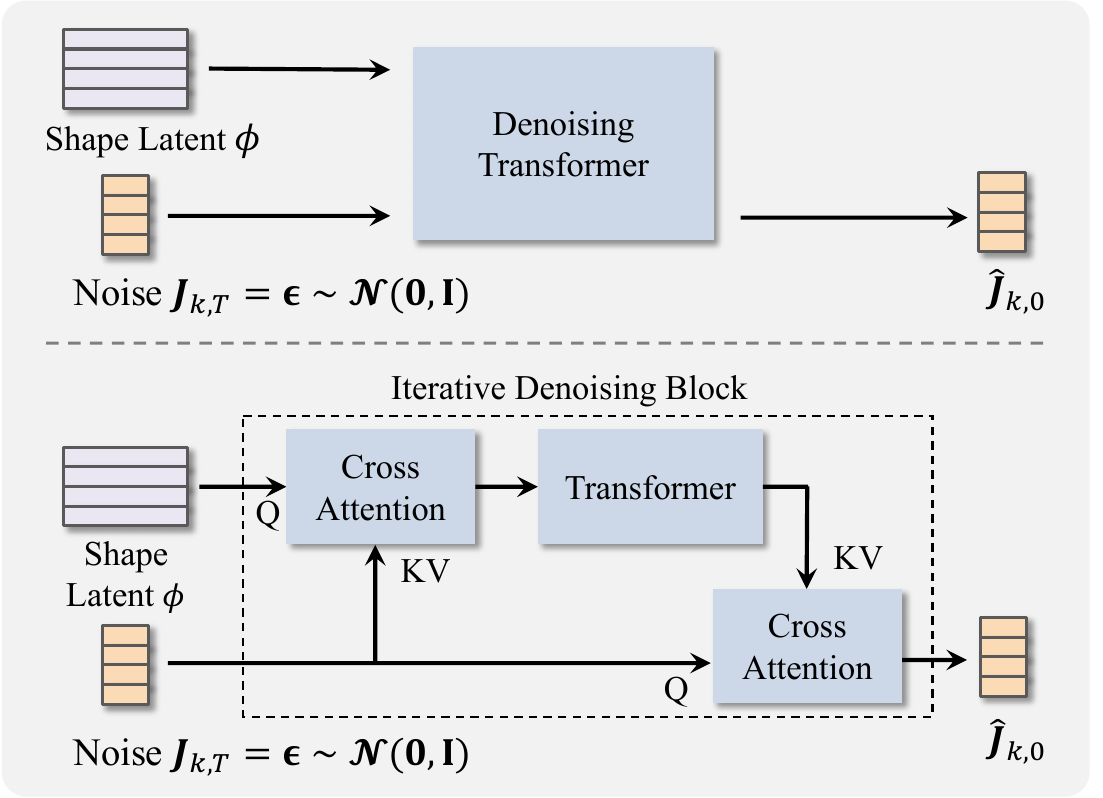}
        \caption{Diffusion-based Architectures}
        \label{subfig:diff}
    \end{subfigure}
    \begin{subfigure}[b]{\linewidth}
        \centering
        \includegraphics[width=0.85\linewidth]{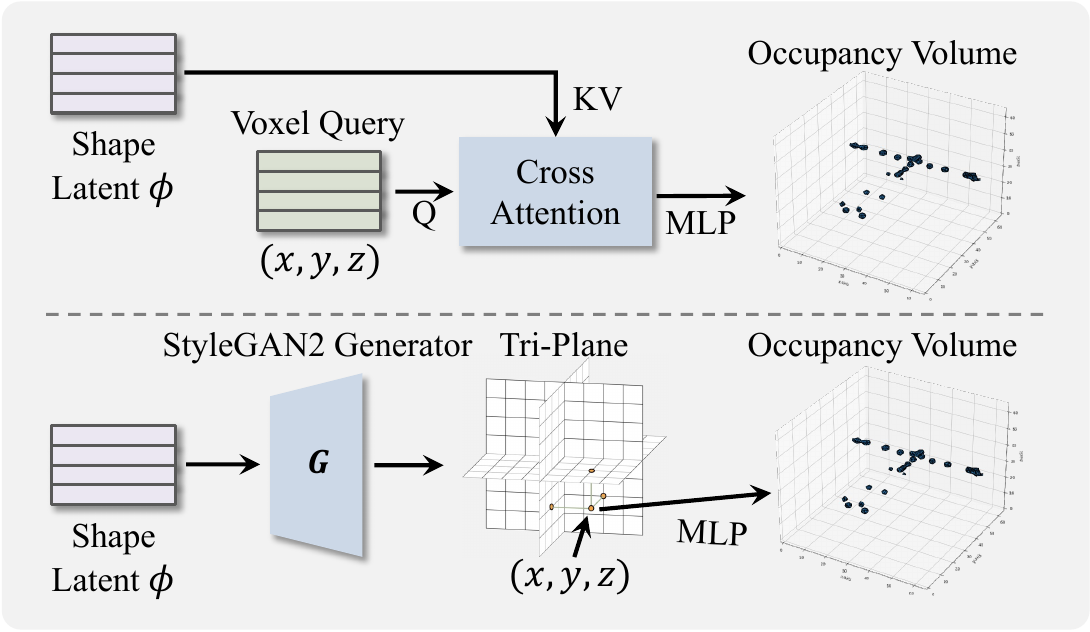}
        \caption{Volume-based Architectures}
        \label{subfig:vol}
    \end{subfigure}
    \vspace{-2em}
    \caption{\textbf{Architectures for Joint Prediction.}
    The model predicts a set of joint locations $\{\hat{\mathbf{J}_k}\}$ from the shape latent $\phi$ (\cref{subsec:joint_arch}).
    }
    \label{fig:joint}
    \vspace{-1.5em}
\end{figure}

\subsubsection{Architectures}
\label{subsec:joint_arch}
We explore three types of architectures for $f_\text{joint}$: regression-based, diffusion-based, and volume-based.

\paragraph{Regression-based.}
\label{meth:regress}
A regression-based model is perhaps the most intuitive approach for joint prediction, which simply regresses a set of 3D coordinates from the extracted shape features and can be directly supervised by the ground-truth joints during training.
However, there are two unique features that make our joint prediction task different from typical regression problems: (1) the number of target joints varies across instances, and (2) the order of the joints is immaterial at this stage\footnote{Although an order could technically be established based on the skeleton, inconsistencies may arise across instances created by different artists.}.
To address these challenges, we design a transformer-based architecture~\cite{jaegle2021perceiver, vaswani2017attention}, which is permutation-equivariant and has proven to be highly scalable, followed by a clustering step~\cite{ester1996density}.

The architecture is illustrated in Fig.~\ref{subfig:regress}.
Specifically, we use a single-layer perceiver-based transformer~\cite{jaegle2021perceiver} with a fixed number of $K_\text{pred}=96$ 
learnable query tokens and the shape latent $\phi$ as key and value tokens, followed by a $3$-layer MLP to produce $K_\text{pred}$ 3D joint locations $\hat{\mathcal{J}}$.
We supervise the model directly using a standard Chamfer Distance between the predicted joints and the ground-truth joints.

Since the number of GT joints ${K_\text{gt}}$ varies across training instances, we cluster the raw predicted joints using the technique of \cite{ester1996density} and use the cluster centers as the final output.

\paragraph{Diffusion-based.}
\label{meth:diff}

Another commonly used architecture for point set generation is diffusion-based models~\cite{sohl2015deep}, which tend to excel with large-scale training.
However, unlike most existing work that focuses either on generating dense point clouds with high geometric redundancy \cite{luo2021diffusion, nichol2022point, huang2024pointinfinity} or on producing a fixed set of semantic keypoints like human poses \cite{Zhou_2023_ICCV,xu20246d}, our model is designed to predict salient joints for animation which may vary across instances.

Following DDPM~\cite{ddpm}, the forward diffusion process injects noise $\bm{\epsilon}_{t} = \{\bm{\epsilon}_{k,t}\}_{k=1}^K$ independently to the GT joints $\mathcal{J}_0 = \{\mathbf{J}_{k,0}\}_{k=1}^{K}$ at a time step $t \in \left[1, T\right]$:
\begin{equation*}
\mathbf{J}_{k,t} = \sqrt[]{\overline{\alpha}_t} \mathbf{J}_{k,0} + \sqrt[]{1-\overline{\alpha}_t} \bm{\epsilon}_{k,t},
\quad \bm{\epsilon}_{k,t} \sim \mathcal{N}(\mathbf{0},\,\mathbf{I}),
\end{equation*}
where $\overline{\alpha}_t$ signifies the noise schedule.
We then train a denoiser $\bm{\epsilon}_{\theta}(\mathcal{J}_{t}, t \mid \phi)$ that learns to predict the noise from the noised joints $\mathcal{J}_t = \{\mathbf{J}_{k,t}\}$ conditioned on the shape latent $\phi$, using the loss:
\begin{equation*}
L_{\text{diff}} = \mathbb{E}_{t, \mathcal{J}_{0}, \bm{\epsilon}_t}
\left[ \| \bm{\epsilon}_{t} - \bm{\epsilon}_{\theta}(\mathcal{J}_{t}, t \mid \phi) \|_2^2 \right].
\end{equation*}

We implement the denoiser using a permutation-equivariant transformer architecture with two variations, as illustrated in Fig.~\ref{subfig:diff}:
\begin{enumerate}
    \item \textbf{Concatenation-based} model inspired by Point-E~\cite{nichol2022point}, where the noisy joints $\mathbf{J}_{k,t}$ are concatenated with the shape latent $\phi$ before passing through a transformer;
    \item \textbf{Cross-attention-based} model inspired by Point-Infinity~\cite{huang2024pointinfinity}, where cross-attention and transformer modules iterate across two branches, between the noisy joints $\mathbf{J}_{k,t}$ and the shape latent $\phi$.
\end{enumerate}
At inference time, we use a DDIM~\cite{ddim} sampler to generate a set of joints $\hat{\mathcal{J}}_0$ from Gaussian noise.
This transformer-based diffusion architecture allows users to specify an arbitrary number of joints at varying levels of granularity during inference by adjusting the shape of the initial noisy point cloud,  as demonstrated in the supp.~mat.

\paragraph{Volume-based.}
\label{meth:volume}
The final option we explore is a dense volume-based model, where the task is framed as predicting the probably of a point in space being near a joint.
To this end, we experiment with two commonly used 3D field representations, an \textbf{implicit neural field} \cite{mildenhall2021nerf} and a \textbf{tri-plane} representation~\cite{chan2022eg3d}, as illustrated in Fig.~\ref{subfig:vol}.

For the implicit neural field, we follow Michelangelo~\cite{zhao2024michelangelo} and implement it using a perceiver-based transformer.
It takes in a set of 3D coordinates $\mathbf{x}_i$ conditioned on the shape latent $\phi$, and predicts a scalar $\sigma(\mathbf{x}_i)$ indicating the probability that a joint is nearby.
For direct supervision, we apply a 3D Gaussian kernel to each GT joint with a standard deviation $\gamma$ ($=2\ \mathrm{voxel}$) and take the maximum response across all joints at each 3D location.
At inference time, we first extract a $64^3$ voxel grid of joint probability scores from the implicit field and apply a threshold to obtain a set of activated grid points.
We then apply the same clustering technique~\cite{ester1996density} to group these points into clusters.

For the tri-plane representation, we follow the implementation of EG3D~\cite{chan2022eg3d} with a StyleGAN2 generator~\cite{Karras2019stylegan2} conditioned on the shape latent.
We then use a similar procedure to extract keypoints from the tri-planes.

\subsubsection{Existing Methods}
We also evaluate two existing auto-rigging methods on our dataset, Pinocchio~\cite{baran2007automatic} and RigNet~\cite{xu2020rignet}.
Pinocchio relies on predefined template skeletons (humanoid or quadruped), and fit a template to a given mesh via optimization.
RigNet proposes a learning-based system similar to ours, but uses primarily Graph Neural Networks.
We adopt the official public implementation for both methods, and train the RigNet model on our larger \dataset Dataset.
We also compare with two commercial auto-rigging tools, Maya~\cite{maya} and Animate Anything~\cite{animateanything}, qualitatively in \cref{subsec:visual}.

\subsubsection{Quantitative Results}

\begin{table}[t]
    \small
    \centering
    \caption{\textbf{Quantitative Evaluation of Joint Prediction.}
    Our proposed architectures scale more effectively with larger training data, outperforming existing methods by a significant margin.
    }
    \begin{tabular}{lccc}
        \toprule
        Model  & CD $\downarrow$ & EMD $\downarrow$ & Training Time\\
        \midrule
        Pinocchio~\cite{baran2007automatic}                     &0.198 & 0.659&-\\
        \midrule
        \multicolumn{4}{l}{\ \ \textit{trained on \datasetacronym-small (14K)}} \\
        RigNet~\cite{xu2020rignet}                       &0.094 & 0.131 &16h\\
        Mich-Regress                  &0.099  & 0.124 &20h\\
        Mich-Diff$_\mathrm{Concat}$ & 0.142& 0.175&26h\\
        Mich-Diff$_\mathrm{Cross}$ & 0.128& 0.160&47h\\
        Mich-Vol$_\mathrm{Implicit}$  & 0.147& 0.127&33h\\
        Mich-Vol$_\mathrm{TriPlane}$  & 0.198& 0.186&20h\\
        Bert-Regress                  & \textbf{0.091}& \textbf{0.113}&11h\\
        Bert-Diff$_\mathrm{Concat}$ & 0.147& 0.177 &65h\\
        Bert-Diff$_\mathrm{Cross}$& 0.137& 0.167&37h\\
        Bert-Vol$_\mathrm{Implicit}$  & 0.123& 0.126&29h\\
        Bert-Vol$_\mathrm{TriPlane}$  & 0.192 &0.171&19h \\
        \midrule
        \multicolumn{3}{l}{\ \ \textit{trained on \datasetacronym-train (225K)}} \\
        RigNet~\cite{xu2020rignet}                       &0.089 & 0.127 &114h \\
        Bert-Regress                  & \textbf{0.077}& \textbf{0.098}& 79h\\
        Mich-Diff$_\mathrm{Cross}$& 0.083& 0.104& 151h\\
        \bottomrule
    \end{tabular}
    \label{tab:joint}
\end{table}

For a comprehensive analysis of different architectures, we first train 10 models on the \datasetacronym-small set, exhausting all combinations of the 2 shape encoders and 5 prediction architectures.
We then evaluate these models and the existing methods (RigNet and Pinocchio) on the \datasetacronym-test set.
For quantitative evaluation, we compute two standard metrics, \textbf{Chamfer Distance} (CD) and \textbf{Earth Mover’s Distance} (EMD), both computed between the predicted and the annotated joints (see supp.~mat.).

The results are summarized in Table~\ref{tab:joint}.
Among the proposed architectures trained on \datasetacronym-small, Bert-Regress outperforms the two existing methods.
We select two architectures for training on the full \datasetacronym-train set: Bert-Regress, and Mich-Diff$_\mathrm{Cross}$.
The regression-based model performs best on the \datasetacronym-small subset, while the diffusion-based model has the potential for scaling with larger training data.
As shown in Table~\ref{tab:joint}, the performance of the model improves as training set increases from 14K to 225K, in particular, from 0.091 to 0.077 for Mich-Diff$_\mathrm{Cross}$.
Despite comparable performance, compared to the regression-based models, the diffusion-based model offers the flexibility for a user to specify an arbitrary number of target joints, but requires a more time-consuming iterative inference procedure.
In comparison, the classic geometric method, Pinocchio~\cite{baran2007automatic}, performs significantly worse.
RigNet~\cite{xu2020rignet} scales poorly with the larger dataset, leading to degraded performance.
We also report the training time in terms of GPU hours on a L40S GPU for  \datasetacronym-small training and on eight TITANRTX GPUs for \datasetacronym-train training.
More analysis is provided in the supp.~mat.

\begin{figure}[t!]
    \centering
    \begin{subfigure}[b]{\linewidth}
        \centering
        \includegraphics[width=\linewidth]{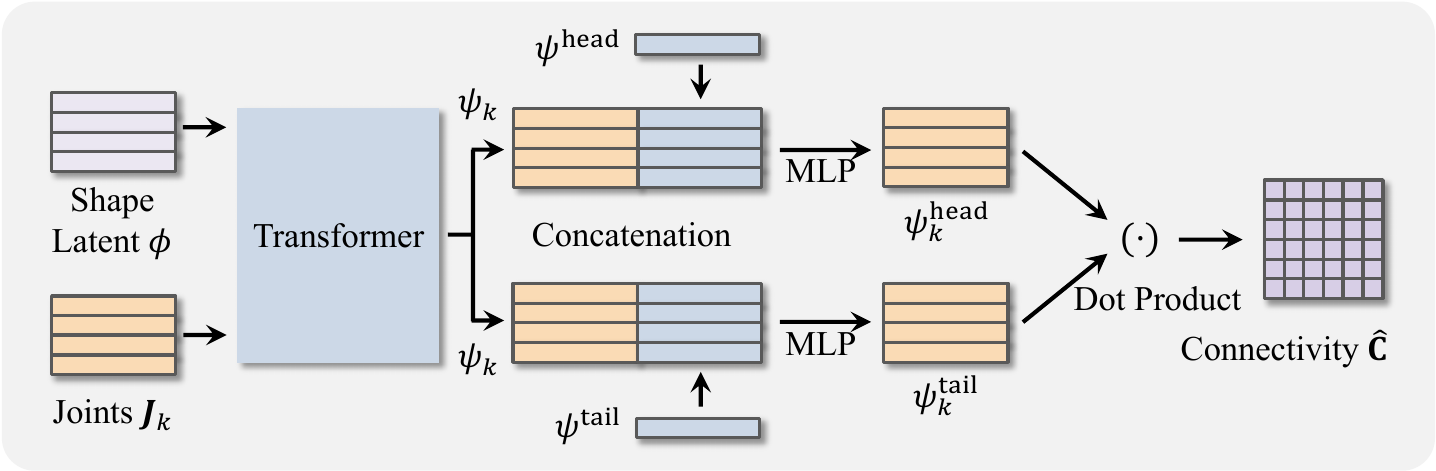}
        \caption{Token-conditioned Architecture}
        \label{subfig:token}
    \end{subfigure}
    \begin{subfigure}[b]{\linewidth}
        \centering
        \includegraphics[width=\linewidth]{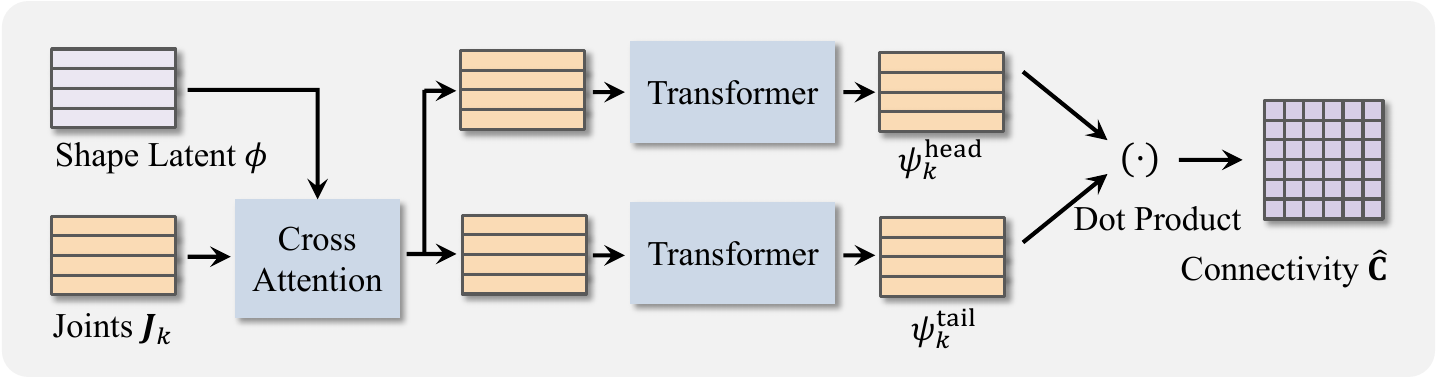}
        \caption{Two-branch Architecture}
        \label{subfig:2branch}
    \end{subfigure}
    \vspace{-2em}
    \caption{\textbf{Architectures for Connectivity Prediction.}
    The model predicts a pair-wise connectivity matrix $\hat{\mathbf{C}}$ from a set of joints $\{\mathbf{J}_k\}$, conditioned on the shape latent $\phi$ (\cref{subsec:connectivity_arch}).
    }
    \label{fig:con}
    \vspace{-1.5em}
\end{figure}

\subsection{Connectivity Prediction}
\label{subsec:connectivity}
After obtaining a set of joints $\{\hat{\mathbf{J}}_k\}_{k=1}^K$, the second module $f_\text{conn}$ predicts connectivity between the joints to establish a bone skeleton.
Specifically, it takes the predicted joints and the shape latent $\phi$ as inputs, and predicts a connectivity matrix $\hat{\mathbf{C}} = f_\text{conn}(\hat{\mathcal{J}}, \phi) \in \mathbb{R}^{K \times K}$, where each element $\hat{c}_{i,j}$ indicates the probability that $\hat{\mathbf{J}}_i$ and $\hat{\mathbf{J}}_j$ are connected by a bone, with $\hat{\mathbf{J}}_i$ as the head joint and $\hat{\mathbf{J}}_j$ as the tail joint.

We train the model using the ground-truth joints $\{\mathbf{J}_k\}_{k=1}^K$ as input and compute binary cross entropy between the predicted and ground-truth  connectivity matrices, $\hat{\mathbf{C}}$ and $\mathbf{C}$.
The GT matrix $\mathbf{C} \in \{0, 1\}^{K \times K}$ can be generated from the annotated bones in the training assets.
During inference, the trained model is used to infer the connectivity of the predicted joints $\{\hat{\mathbf{J}}_k\}_{k=1}^K$.

In practice, Blender only allows each bone to have \emph{at most one} parent bone.
This implies that each joint can only serve as the \emph{tail} joint of either one single bone or none (\ie, a root joint), meaning the sum of each column of the GT connectivity matrix $\mathbf{C}$ is either 1 or 0 (root).
Therefore, if a joint $\mathbf{J}_i$ is a root joint, we assign the $(i,i)$ entry of $\mathbf{C}$ to 1, \ie,  $c_{i,i}=1$, while keeping the rest of the column as 0.
At inference time, we can exploit this fact and simply take the \texttt{argmax} of each column of the predicted connectivity matrix $\hat{\mathbf{C}}$ to identify root joints or retrieve the head joints.
This enables us to construct a kinematic chain for animation.

\subsubsection{Architectures}
\label{subsec:connectivity_arch}
We design two architectures to predict this connectivity matrix $\hat{\mathbf{C}}$ given a set of joints $\{\mathbf{J}_k\}_{k=1}^K$ conditioned on the shape latent $\phi$: a \textbf{token-conditioned} architecture and a \textbf{two-branch} architecture, as illustrated in Fig.~\ref{fig:con}.

\paragraph{Token-conditioned.}
Given the input joint coordinates, this model first applies positional encoding and extracts embeddings via a linear layer.
These joint embeddings together with the shape latent $\phi$ go through a transformer to obtain a set of per-joint feature tokens $\{\psi_k\}$.
Each token $\psi_k$ is then concatenated with a (shared) learnable `head' token $\psi^{\text{head}}$ and a (shared) learnable `tail' token $\psi^{\text{tail}}$, and goes through two separate MLPs, generating a pair of head-conditioned and tail-conditioned tokens $\psi_k^{\text{head}}$ and $\psi_k^{\text{tail}}$.
Finally, we simply compute the dot product between two sets of tokens $\{\psi_k^{\text{head}}\}$ and $\{\psi_k^{\text{tail}}\}$ to obtain the output matrix $\hat{\mathbf{C}}$.

\paragraph{Two-branch.}
Instead of conditioning on `head' and `tail' tokens, 
The two-branch model processes the joint embeddings through two independent transformer branches, after incorporating the shape latent via cross attention.
Similarly, the final matrix $\hat{\mathbf{C}}$ is obtained by taking the dot product between the two sets of output tokens from each branch.

\subsubsection{Existing Methods}
We compare our models with the bone prediction module of RigNet~\cite{xu2020rignet} on our dataset.
Rather than using the dot product, they directly feed in each pair of joints to an MLP and predict their connectivity, which is less computationally efficient ($O(K^2)$).
Furthermore, they use a separate network to predict the probability of a joint being a root.
In comparison, our models are more efficient by directly obtaining a connectivity matrix via a dot product.

\begin{table}[t]
    \small
    \centering
    \caption{\textbf{Quantitative Evaluation of Connectivity Prediction.}
    Despite overfitting on the smaller training set, our proposed architectures scale effectively with larger training data.
    }
    \begin{tabular}{lccc}
        \toprule
        Model   & Precision $\uparrow$ & Recall $\uparrow$& Training Time\\
        \midrule
        \multicolumn{4}{l}{\ \ \textit{trained on \datasetacronym-small (14K)}} \\
        RigNet~\cite{xu2020rignet}        & 60.6\%& 57.9\%& 8h\\
        Mich-Token        &58,0\% &55.8 \% & 21h\\
        Mich-2Branch    & 60.5\%& 58.5\% & 19h \\
        Bert-Token        & 59.3\% & 57.2 \%& 15h\\
        Bert-2Branch    & \textbf{61.6\%}& \textbf{59.7\%}& 12h\\
        \midrule
        \multicolumn{4}{l}{\ \ \textit{trained on \datasetacronym-train (225K)}} \\
        RigNet~\cite{xu2020rignet}       &47.9\% &50.4\%& 51h\\
        Bert-2Branch    & \textbf{84.6\%}& \textbf{83.5\%}& 76h\\
        \bottomrule
    \end{tabular}
    \label{tab:con}
\end{table}

\subsubsection{Quantitative Results}
Similar to joint evaluation, we evaluate all architecture combinations on \datasetacronym-small, and train the top performing models on the full \datasetacronym-train set.
We report two standard metrics for comparison: 
\textbf{Precision} and \textbf{Recall}.

The results are summarized in Table~\ref{tab:con}.
On \datasetacronym-small, we in fact observe overfitting for all of our models, whereas RigNet performs reasonably well.
However, on the full \datasetacronym-train set, we observe the opposite: RigNet severely underfits potentially due to a limited model size, whereas our proposed architectures scale well with the larger dataset, outperforming RigNet by a significant margin.
Additionally, the two-branch model tends to perform better than the token-conditioned model.

\subsection{Skinning Weight Prediction}
\label{subsec:skin}
After obtaining the bones $\mathcal{B}$, the final step is to predict the skinning weights $\hat{\mathbf{W}}_i = f_\text{skin}(\mathbf{V}_i \mid \mathcal{B}, \phi) \in \mathbb{R}^B$ of each vertex $\mathbf{V}_i$ with respect to the bones, where $B=|\mathcal{B}|$ denotes the number of bones.
In practice, our model takes in the joint locations $\mathbf{J}$ of the bones, and for efficiency, processes a set of points in a batch using a transformer architecture, producing a skinning weight matrix $\hat{\mathbf{W}} \in \mathbb{R}^{|\mathbf{V}| \times B}$.
For training, we use the resampled point clouds explained in \cref{subset:preprocessing} together with the annotated bones as inputs, and compute the cosine similarity loss against the ground-truth skinning weights $\mathbf{W}$ (see supp.~mat. for details).
During inference, we can infer the skinning weights of the original mesh vertices over the predicted bones.

\begin{figure}[t!]
    \centering
    \begin{subfigure}[b]{\linewidth}
        \centering
        \includegraphics[width=\linewidth]{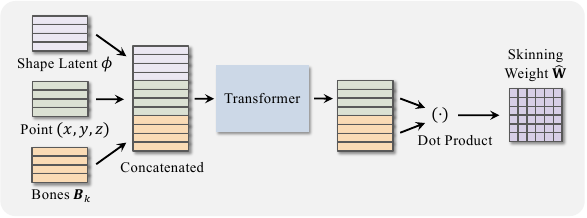}
        \caption{Concatenation-based Architecture}
        \label{subfig:concat}
    \end{subfigure}
    \begin{subfigure}[b]{\linewidth}
        \centering
        \includegraphics[width=\linewidth]{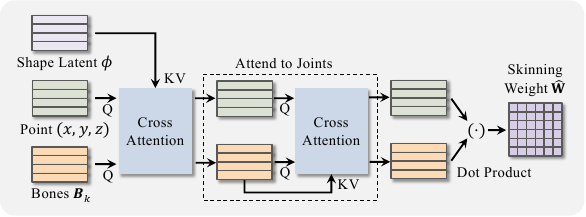}
        \caption{Cross-attention-based Architecture}
        \label{subfig:cross}
    \end{subfigure}
    \vspace{-2em}
    \caption{\textbf{Skinning Weight Prediction Architectures.}
    The model predicts a pair-wise skinning weight matrix $\hat{\mathbf{W}}$ between each point and each bone $\mathbf{B}_k$, conditioned on the shape latent $\phi$ (\cref{subsec:skin_arch}).
    }
    \label{fig:skin}
    \vspace{-1.5em}
\end{figure}

\subsubsection{Architectures}
\label{subsec:skin_arch}
We experiment with two architectures illustrated in Fig.~\ref{fig:skin}.

\paragraph{Concatenation-based.}
This model simply concatenates the point embeddings with the bone embeddings and shape latent $\phi$, and mixes these embeddings to generate two set of token features: per-point tokens $\{\phi^\text{point}_i\}_{i=1}^{|\mathbf{V}|}$ and per-bone tokens $\{\phi^\text{bone}_b\}_{b=1}^B$.
Reusing the dot product trick in connectivity prediction followed by a \texttt{softmax} over all bones, we obtain a skinning weight matrix $\hat{\mathbf{W}}$.

\paragraph{Cross-attention-based.}
The second variant stacks two consecutive cross-attention layers. The per-point tokens $\{\phi^\text{point}_i\}_{i=1}^{|\mathbf{V}|}$ and per-bone tokens $\{\phi^\text{bone}_b\}_{b=1}^B$ is generated by passing the cross-attention layers.
Finally, the skinning weight matrix $\hat{\mathbf{W}}$ is obtained via a dot product followed by a \texttt{softmax}.

\begin{figure*}[t]
    \centering
    \includegraphics[width=\linewidth]{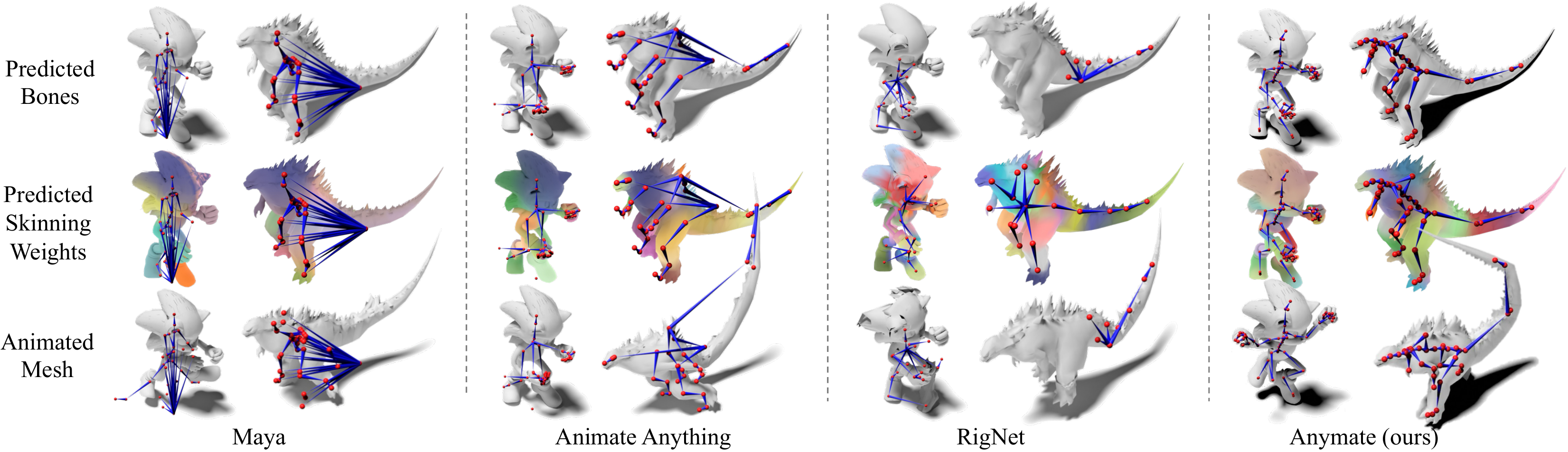}
    \vspace{-2em}
    \caption{\textbf{Visual Results and Comparisons with Existing Methods.}
    All methods estimate a bone skeleton and skinning weights from an input 3D mesh.
    Commercial auto-rigging tools like Maya~\cite{maya} and Animate Anything~\cite{animateanything} typically rely on template skeletons and generalize poorly well to complex shapes.
    Despite training on our larger dataset, RigNet~\cite{xu2020rignet} also struggles to predict reasonable skeleton and skinning weights, resulting in severe distortion in animation.
    In contrast, our model produces more plausible predictions and animations.
    }
    \label{fig:result}
    \vspace{0 em}
\end{figure*}

\subsubsection{Existing Methods}
We compare against RigNet~\cite{xu2020rignet} as well as a traditional geometric method, GeoVoxel~\cite{dionne2013geodesic}.
RigNet uses a graph neural network based architecture and only predicts the skinning weights with respect to the five nearest bones.
In practice, we found this often insufficient, in particular with erroneous joint predictions.
We train and evaluate RigNet on our proposed dataset using the official implementation and the same hyper-parameters. 
GeoVoxel calculates skinning weights simply based on the geodesic distance after voxelization.
We evaluate GeoVoxel on our test set using the implementation from Maya~\cite{maya}.

\begin{table}[t]
    \small
    \centering
    \caption{\textbf{Quantitative Evaluation of Skinning Weight Prediction.}
    Our proposed architectures scale effectively with larger training data, outperforming GeoVoxel~\cite{dionne2013geodesic} and RigNet~\cite{xu2020rignet}.
    }
    \begin{tabular}{lcccc}
        \toprule
        Model    & CE $\downarrow$ & Cos $\uparrow$ & MAE $\downarrow$ & Train Time  \\
        \midrule
        GeoVoxel                   & 3.569& 0.407& 0.056& -\\
        \midrule
        \multicolumn{4}{l}{\ \ \textit{trained on \datasetacronym-small (14K)}} \\
        RigNet                     &1.573 &0.671 &0.031 &11h    \\

        Mich-Concat                    & 1.169&  0.754&  0.029&  41h \\
        Mich-Cross                     & 1.270& 0.766 & 0.026 &  21h\\

        Bert-Concat                      & 1.148& 0.768 & 0.027& 53h   \\
        Bert-Cross                       & \textbf{1.134}& \textbf{0.801}& \textbf{0.025}& 32h \\
        \midrule
        \multicolumn{4}{l}{\ \ \textit{trained on \datasetacronym-train (225K)}} \\
        RigNet                  &1.521 &0.693 & 0.028& 49h    \\

        Bert-Cross                & \textbf{0.741} & \textbf{0.915}& \textbf{0.014} &  83h \\ 
        \bottomrule
    \end{tabular}
    \label{tab:skin}
\end{table}

\subsubsection{Quantitative Results}
Similarly, we train the models on both the \datasetacronym-small and \datasetacronym-train sets, and evaluate on the \datasetacronym-test set.
We report three metrics: \textbf{Cross Entropy} (CE), \textbf{Cosine Similarity} (Cos), and \textbf{Mean Absolute Error} (MAE).

The results are summarized in Table~\ref{tab:skin}.
Overall, GeoVoxel performs poorly on our test set, serving as a purely geometry-based lower-bound.
Our proposed architectures scale well with the larger training set, outperforming the baselines by a significant margin.
In contrast, little improvement is observed with RigNet as the training data size increases.
Among our proposed architectures, Bert-Cross achieves the best results.

\begin{figure*}[t]
    \centering
    \includegraphics[width=0.85\linewidth]{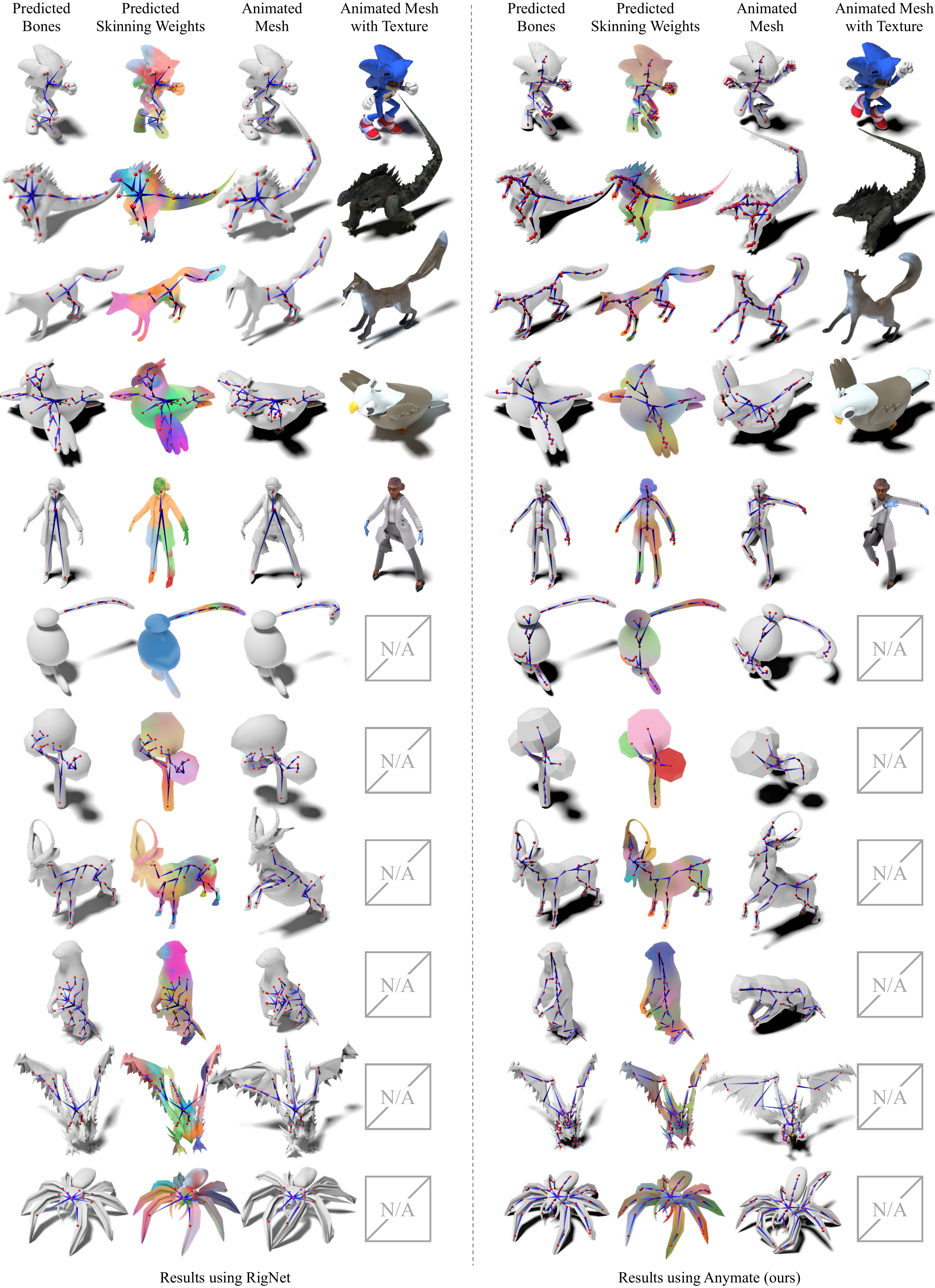}
    \caption{\textbf{Additional Qualitative Comparison with RigNet~\cite{xu2020rignet}.}
    Our model consistently produces more accurate predictions.
    Note that some of the assets do not come with texture, hence marked with `N/A'.
    }
    \label{fig:more-results}
\end{figure*}

\subsection{Visual Results}
\label{subsec:visual}
After all individual modules have been trained, we can run inference on any new test mesh and obtain the skeleton and skinning weights automatically in less than a minute.
With these predictions, we can easily animate the object by manipulating the predicted skeleton, as visualized in Fig.~\ref{fig:more-results} and Fig.~\ref{fig:result} .
For all visualizations, we use the best-performing model variants for inference, namely, Mich-Regress (joint), Bert-2Branch (connectivity), and Bert-Cross (skinning).

We compare our results against the previous method, RigNet~\cite{xu2020rignet}, and two commercial auto-rigging tools, Maya~\cite{maya} and Animate Anything \cite{animateanything}.
Maya only supports fitting a humanoid skeleton to the input mesh and generalizes poorly to non-humanoid objects.
Animate Anything, on the other hand, classifies the object into one of a few categories and fits a category-specific skeleton.
Both tools produce inaccurate skeletons for complex geometries, resulting in unrealistic animations.
For a fair comparison, we retrain the RigNet model on our full dataset.
However, it still generalizes poorly to challenging instances.
In comparison, our model predicts highly accurate bone skeleton and skinning weights, enabling realistic animations.

\section{Conclusions}
We present the \dataset Dataset, the largest object rigging dataset to date, containing 230K 3D assets with rigging and skinning information.
Using this dataset, we proposed an effective learning-based auto-rigging framework, with three modules for joint, connectivity, and skinning prediction.
For each module, we carefully designed a set of baselines and thoroughly evaluated their performance using the proposed dataset.
Experiment results have demonstrated that our proposed baselines scale effectively with the larger training data, outperforming previous methods and existing commercial auto-rigging tools, and provide a reference point for future research in data-driven auto-rigging.

\begin{acks}
This work is in part supported by ONR YIP N00014-24-1-2117, NSF RI \#2211258 and \#2338203, the Stanford Institute for Human-Centered AI (HAI), the Stanford Center for Integrated Facility Engineering (CIFE), and Meta. 
\end{acks}

\clearpage
\bibliographystyle{ACM-Reference-Format}

\bibliography{ref}%

\input{sec/6_supmat}
\end{document}

%% file: sec/6_supmat.tex
\renewcommand{\thesection}{{S.}\arabic{section}}
\renewcommand{\thefigure}{{S}\arabic{figure}}
\renewcommand{\thetable}{{S}\arabic{table}}
\renewcommand{\theequation}{{S}\arabic{equation}}

\clearpage
\setcounter{page}{1}
\appendix

\input{fig/bones}
\section{Additional Details in Data Processing}
\subsection{Augmentation}
To enhance the diversity, we augment \dataset Dataset by incorporating key frame poses from the original animations. Specifically, we first divide \dataset Dataset into a training set (93K samples) and a test set (5.6K samples). For each animated asset in the training set, we then randomly select 1–3 key frames that exhibit notable pose variations and add the corresponding meshes and skeletons in these new poses to the dataset. This augmentation enhances pose diversity, enabling the model to learn skeletal representations across varying configurations better.

\subsection{Processing}
In some cases, when a bone is never used (with no associated skinning weights), we simply remove it and assign its children bones directly to its parent bone.
If the asset is a scene, isolated mesh parts with no skinning weights are removed, and the rest are merged.
We then center the mesh at the origin and scale it to fit within $[-1, 1]$.
We also correct the bones by fixing the noisy tail joints' position as illustrated in \cref{fig:bones}.

\subsection{Manual inspection for quality assurance}
To assess the quality of our processed assets, we manually inspected a subset of 5000 random samples.
We found 4.6\% of them to be suboptimal (mainly due to the issues with annotated skeletons), which indicates our processing pipeline is quite effective.
Moreover, we removed all such low-quality assets in the test set to ensure a reliable benchmark.

\section{Additional Results}

\input{supmat/fig/fig_animation}
\subsection{Additional Qualitative Results}
We provide 3D visualizations of the animation results in the \textbf{website} enclosed in the supplementary material.
Animation results can be seen on the \textbf{website} and \cref{fig:animation}.

To demonstrate the generalization capability of our models, we additionally test our model on novel 3D meshes generated by a recent 3D generation method~\cite{tochilkin2024triposr}.
The predicted rig and animation results are included in \cref{fig:generated} and in the website.

\subsection{Sampling Varying Numbers of Joints from the Diffusion-based Architectures}

As mentioned in Sec.~4.2.1 of the main paper, the diffusion-based architectures support inference with an arbitrary number of predicted joints specified by a user.
\cref{fig:diff-vary} shows an example with two different number of joints, 32 and 64.
Both predicted skeletons look reasonable based on the input mesh.
An interesting observation is that with an increased number of joints from 32 to 64, the model tends to predict additional joints around the hands, while maintaining the sparsity at other locations, \eg single joints at the knees.
This offers an intuitive control over the granularity of the predicted rigs.

In addition, we also find an extra non-maximum suppression (NMS) step leads to cleaner results and hence better numbers on our benchmarks, as visualized on the last column in \cref{fig:diff-vary}.
Therefore, for all the quantitative evaluations of the diffusion-based models, we use this extra NMS step.
Specifically, we sample 128 points and retain only one joint with each $0.05^3$ voxel.

\input{supmat/fig/fig_diff}
\input{supmat/fig/fig_generated}

\subsection{Additional Quantitative Results}
\input{supmat/tab/result_joint_supp}
\input{supmat/tab/result_skin_supp}
\input{supmat/tab/result_joint_RigNetTest}
\input{supmat/tab/result_con_RigNetTest}
\input{supmat/tab/result_skin_RigNetTest}

\input{tab/Joints_RigNet}

\input{tab/Con_RigNet}

\input{tab/Skeleton_RigNet}

\input{tab/Skin_RigNet}
We include four additional sets of evaluations.
First, we additionally report the Precision and Recall metrics for both joint and skinning predictions, as explained in \cref{sec:supp_tech}.
Second, we also evaluate the performance of RigNet~\cite{xu2020rignet} using the pre-trained weights officially released by the authors, as an additional baseline, marked with $^\dagger$ in all tables.
What's more, we evaluate all of the models on the same test set used by RigNet~\cite{xu2020rignet} (RigNet-test) containing 270 instances, and report the results in \cref{tab:supp_joint_rignet,tab:supp_con_rignet,tab:supp_skin_rignet}.
Lastly, for a more comprehensive comparison of the models' efficiency, we conducted training of our proposed baselines on the RigNet-train dataset and evaluated their performance on the RigNet-test dataset. The results are reported in \cref{tab:rubb_joint_rignet,tab:rubb_con_rignet,tab:rubb_skel_rignet,tab:rubb_skin_rignet}. 
We found that our model consistently outperforms RigNet significantly, even on the RigNet-train set. The key factor behind this enhancement is our elimination of certain assumptions present in RigNet, combined with our shift from GNN-based to transformer-based architectures for rigging prediction.

For joint prediction, from \cref{tab:supp_joint,tab:supp_joint_rignet}, it is evident that all our models trained on the \dataset-train exhibit superior performance, compared to the existing methods across all metrics.
On RigNet-test, the RigNet model trained on our \dataset Dataset in fact performs worse compared to the originally released weights (pretrained on the RigNet dataset).
This could be attributed to the fact that the training data used for the released weights model aligns closely with the domain of the RigNet-test.

For connectivity prediction, based on \cref{tab:supp_con_rignet}, we can observe that the pre-trained RigNet has a competitive performance compared to our models trained on \dataset-small. However, when trained on \dataset-train, our models greatly surpass the performance of the pre-trained RigNet.
This emphasizes the scalability of our framework.

For skinning weight prediction, based on \cref{tab:supp_skin,tab:supp_skin_rignet}, we can observe that RigNet demonstrates notably high recall but low precision, primarily because it predicts the skinning weight for only the five nearest bones.
This limitation often leads to weights being assigned to all five nearest bones, which in turn affects precision negatively.
Conversely, our model strikes a good balance between precision and recall, performing the best on both the \dataset-test and RigNet-test.

\section{Inference Time Analysis}

We have reported the training time of different models in the tables in the main paper.
During inference, RigNet takes over 10 minutes per instance to predict skeleton and skinning weights, while our final model only takes less than 3 seconds, significantly enhancing efficiency.

\section{Additional Technical Details}
\label{sec:supp_tech}
\subsection{Joint Prediction}
\paragraph{Metrics.}
As mentioned in Sec.~4.2.3 in the main paper, we use two metrics for quantitative evaluation of the joint prediction accuracy: Chamfer Distance (CD) and Earth Mover's Distance (EMD).
The Chamfer Distance is computed as follows:
\begin{equation}
\begin{aligned}
    L_\text{CD}(\hat{\mathcal{J}}, \mathcal{J}) &= \frac{1}{K_\text{pred}} \sum_{\hat{\mathbf{J}}_i \in \hat{\mathcal{J}}} \min_{\mathbf{J}_j \in \mathcal{J}} \| \hat{\mathbf{J}}_i - \mathbf{J}_j \|_2^2 \\
    &+ \frac{1}{K_\text{gt}} \sum_{\mathbf{J}_j \in \mathcal{J}} \min_{\hat{\mathbf{J}}_i \in \hat{\mathcal{J}}} \| \mathbf{J}_j - \hat{\mathbf{J}}_i \|_2^2,
\end{aligned}
\end{equation}
where $\hat{\mathcal{J}} = \{\hat{\mathbf{J}}_i\}_{i=1}^{K_\text{pred}}$ and $\mathcal{J} = \{\mathbf{J}_j\}_{j=1}^{K_\text{gt}}$ are the two sets of predicted and (annotated) ground-truth (GT) 3D joint locations respectively.

The Earth Mover's Distance is defined as solving the optimization problem of:
\begin{equation}
    L_\text{EMD}(\hat{\mathcal{J}},\mathcal{J})=\min_{f_{ij}}
    \frac{1}{K_\text{pred}}\sum_{i=1}^{K_\text{pred}}\sum_{j=1}^{K_\text{gt}}d_{ij}\cdot f_{ij},
\end{equation}
where $d_{ij}= \| \hat{\mathbf{J}}_i-\mathbf{J}_j \|_2^2 $ is the distance between two joints, $f_{ij}>0$ is the flow that satisfies following:
\begin{equation}
\sum_{i=1}^{K_\text{pred}}{f_{ij}}=\frac{1}{K_\text{gt}},
~~ 
\sum_{j=1}^{{K_\text{gt}}}{f_{ij}}=\frac{1}{K_\text{pred}},
~~
\sum_{i=1}^{K_\text{pred}}\sum_{j=1}^{{K_\text{gt}}}{f_{ij}}=1.
\end{equation}
In practice, we use the implementation in Point-Cloud-Utils~\cite{pcu} to calculate the EMD.

Both metrics evaluate the ``closeness'' between the two sets of points.
In addition, we also evaluate the ``correctness'' of the predicted joints using the standard Precision metric, and the ``coverage'' using the Recall metric, after identifying the closest mapping between two point sets.
Specifically, Precision is calculated as the ratio of correctly predicted joints (those with a distance to their nearest GT joint lower than the threshold) to the total number of predicted joints.
Recall, on the other hand, is the ratio of correctly matched GT joints to the total number of GT joints.
The threshold is set to 0.05 here.

\subsection{Connectivity Prediction}
\paragraph{Metrics.}
Similarly, we use Precision and Recall to assess the connectivity prediction.
Precision is computed as the proportion of correctly predicted pairs among all predicted pairs, whereas Recall is the proportion of correctly predicted pairs among all GT pairs.

\subsection{Skinning Weight Prediction}
\paragraph{Loss Function.}
As described in Sec.~4.4, we train the skinning weight prediction models using the cosine similarity loss.
Specifically, given the two predicted and ground-truth skinning weight matrices, 
$\hat{\mathbf{W}}$ and $\mathbf{W}$, across $N$ vertex and $K_\text{gt}$ joints, the loss is computed as:
\begin{equation}
    L_\text{cos}(\hat{\mathbf{W}},\mathbf{W}) 
= \frac{1}{N} \sum_{i=1}^{N} \left( 1- \texttt{cos}(\mathbf{\hat{W}}_i, \mathbf{W}_{i}) \right),
\end{equation}
where \texttt{cos} denotes the cosine similarity between the two vectors $\mathbf{\hat{W}}_i$ and $\mathbf{W}_{i}$.

\paragraph{Metrics.}
We report three evaluation metrics for the skinning weight prediction: Cross Entropy (CE), Cosine Similarity (Cos), and Mean Absolute Error (MAE).

MAE is computed as the average difference between each vertex and each bone:
\begin{equation}
    L_\text{MAE}(\hat{\mathbf{W}},\mathbf{W}) = \frac{1}{N {K_\text{gt}}} \sum_{i=1}^{N} \sum_{j=1}^{K_\text{gt}} | \hat{w}_{ij} - w_{ij} |,
\end{equation}
where $\hat{w}_{ij}$ and $w_{ij}$ are the elements of the predicted and GT skinning weight matrices, $\hat{\mathbf{W}}$ and $\mathbf{W}$, respectively.

CE is computed as the average cross-entropy over all vertices with the ground truth: 
\begin{equation}
   L_\text{CE}(\hat{\mathbf{W}},\mathbf{W}) = \frac{1}{N {K_\text{gt}}} \sum_{i=1}^{N} \sum^{K_\text{gt}}_{j=1} \left( - w_{ij} \log {{\hat{w}_{ij}}} \right).
\end{equation}

In addition, we also report the Precision and Recall in \cref{tab:supp_skin,tab:supp_skin_rignet}.
For skinning prediction, Precision is determined by identifying the set of bones that significantly ``influence'' each vertex, with a skinning weight exceeding a threshold of 0.05.
It is calculated as the proportion of the ``influential'' bones identified based on the predicted skinning that align with those identified based on the ground-truth skinning weights.

Recall, conversely, is computed as the ratio of ``influential'' bones identified based on the ground-truth skinning weights that successfully match those derived from the predicted skinning weights.

\section{Rest Pose Derivation}
As mentioned in Sec.~3.1 of the main paper, the rest pose transformation $G_b(\xi^*)$ can be derived from the head and tail joint locations of the bones, $\{(\mathbf{J}_b^\text{head}, \mathbf{J}_b^\text{tail})\}_{b=1}^B$, in a kinematic tree structure.

Specifically, $G_b(\xi^*)$ (in the world coordinate frame) is a composition of the transformation $g_b$ of the bone $b$ itself (in its local coordinate frame) and the composed transformation $G_{\pi(b)}$ of its parent, bone $\pi(b)$, following the kinematic tree:
\begin{equation}
    G_b = G_{\pi(b)} \circ g_b,
\end{equation}
where $G_1 = g_1$ and $g_b(\xi_b^*) \in SE(3)$ is a $4 \times 4$ transformation matrix, with a rotation matrix $\mathbf{R}_{\xi_b^*} \in \mathbb{R}^{3 \times 3}$ and a translation vector $\mathbf{t}_{\xi_b^*} \in \mathbb{R}^{3 \times 1}$:
\begin{equation}
    g_b(\xi_b^*) = \begin{bmatrix}
        \mathbf{R}_{\xi_b^*} & \mathbf{t}_{\xi_b^*} \\ 0 & 1 \\
    \end{bmatrix}.
\end{equation}

Note that the local coordinate frame of each bone $b$ can be defined arbitrarily with $\mathbf{R}_{\xi_b^*}$ and $\mathbf{t}_{\xi_b^*}$.
In particular, we choose an intuitive definition, where each bone is a unit vector originating from the origin pointing towards $(0, 0, 1)$ on the $z$-axis of its local coordinate frame, and the $x$-axis is always horizontal (\wrt the world coordinate frame).
As such, the rest-pose rotation matrix $\mathbf{R}_{\xi_b^*}$ and translation vector $\mathbf{t}_{\xi_b^*} = (0, 0, |\mathbf{J}_b^\text{tail} - \mathbf{J}_b^\text{head}|)$ can be obtained via a similar procedure as the camera lookat matrix (from head joint $\mathbf{J}_b^\text{head}$ to tail joint $\mathbf{J}_b^\text{tail}$).

\input{tab/implementation}
\section{Implementation Details}
Details of the architecture designs for each module and hyper-parameter setting are listed in
\cref{tab:implementation}.

\input{supmat/fig/fig_failure}

\section{Limitation and Failure Cases Analysis}
While our framework has proven efficient and robust on a wide range of examples, we do observe some limitations that require further exploration in future work.
Firstly, despite training on a vast dataset, our model occasionally makes incorrect skeleton predictions under certain circumstances. There are two common failure cases: (1) when some parts of the mesh penetrate or are too close to other parts, it may cause failure in skinning prediction, as illustrated in \cref{fig:failure}; (2) when the mesh contains some bulky parts, it may predict too many joints within those regions, resulting in issues in the subsequent modules, as depicted in \cref{fig:failure}. (1) could be addressed by incorporating geometric regularizations and/or pre-processing the input mesh, whereas (2) could potentially be mitigated with an adjustable granularity control. In addition, skeleton-based rigging systems face inherent limitations in modeling certain complex deformations, such as simulating elastic materials or capturing nuanced facial expressions. These challenges may be better addressed through alternative rigging representations like neural cages or other physics-informed approaches. We leave the exploration of these specialized deformation problems to future work.

%% file: fig/bones.tex
\begin{figure}[t]
    \centering
    \begin{subfigure}{0.45\linewidth}
        \centering
        \includegraphics[width=\linewidth]{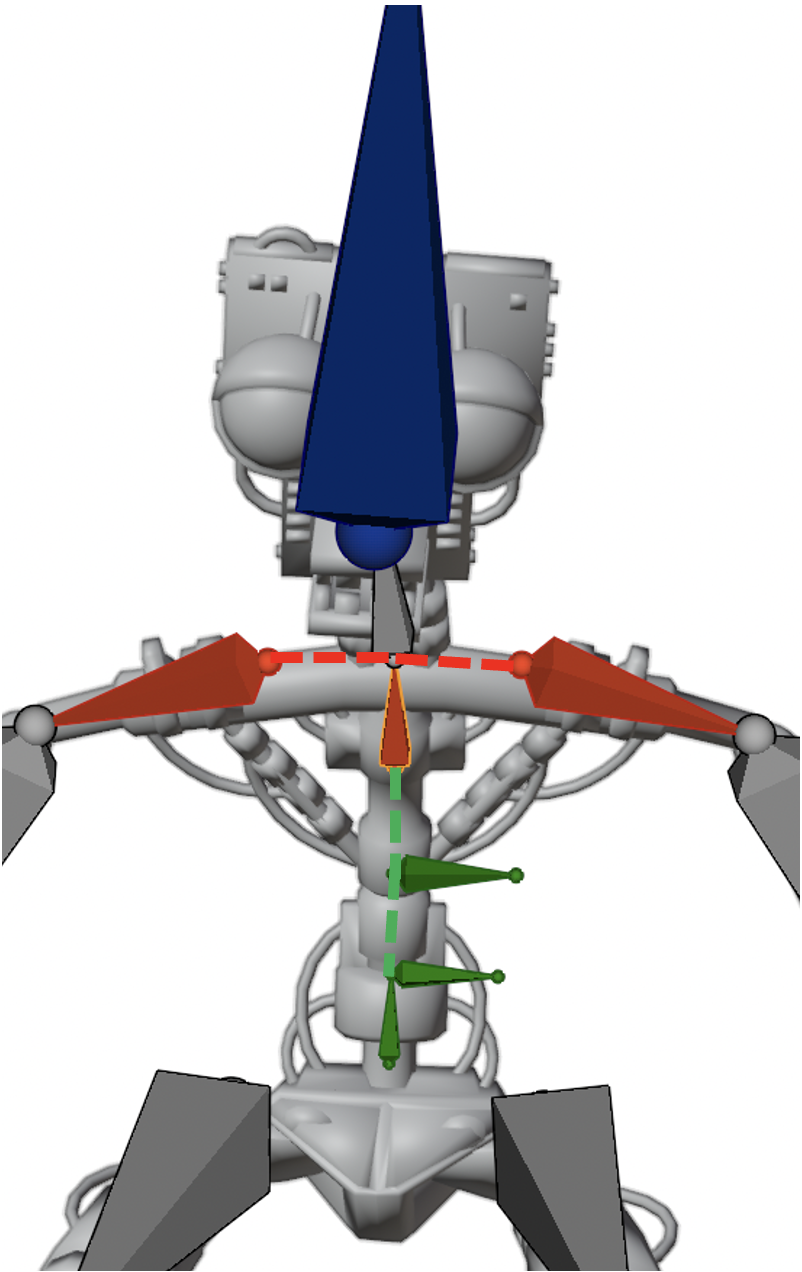}
        \caption{Original Bones}
        \label{subfig:origin}
    \end{subfigure}
    \hfill
    \begin{subfigure}{0.45\linewidth}
        \centering
        \includegraphics[width=\linewidth]{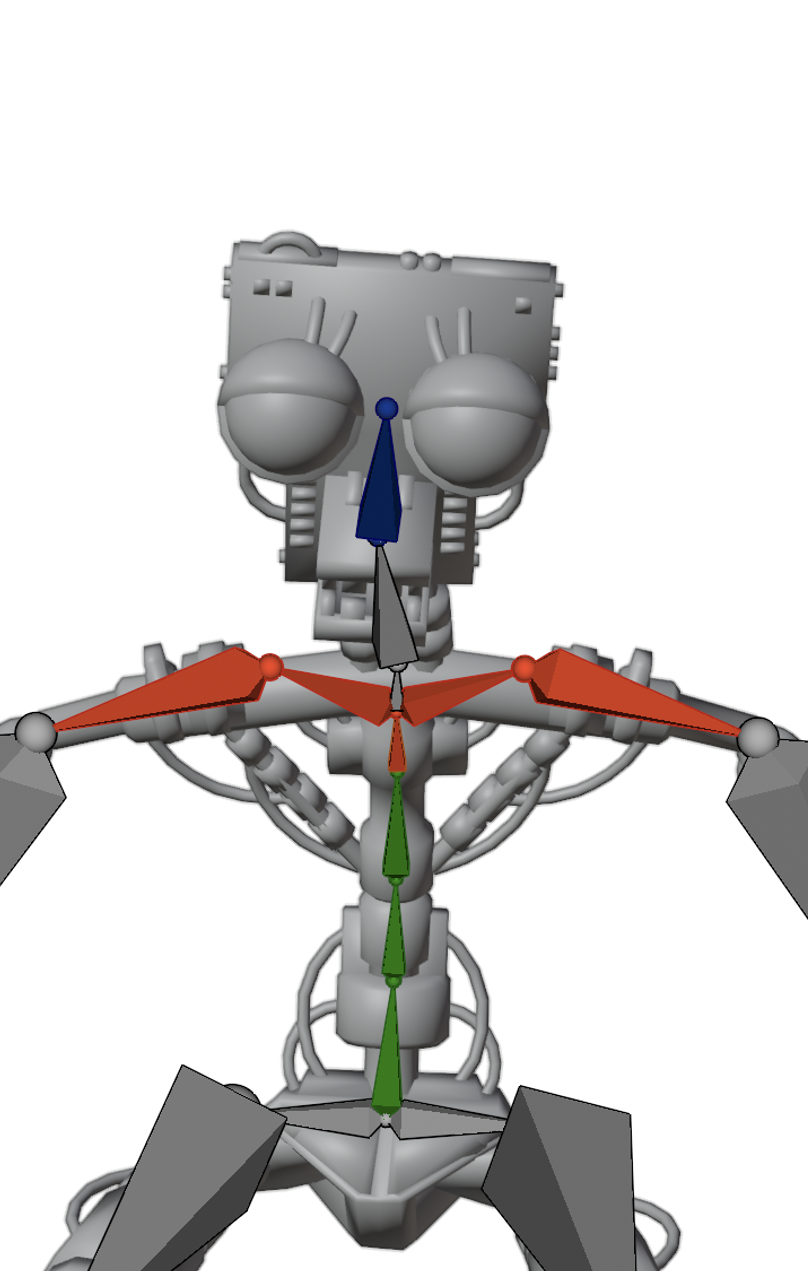}
        \caption{Fixed Bones}
        \label{subfig:fixed}
    \end{subfigure}
    \vspace{-0.8em}
    \caption{
    \textbf{Bones Correction.} Dashed lines are parent-ship in the original data. Tail position is assigned with the child's position (Green case). When a bone has multiple children, tail position is assigned with skinning direction, and extra bones are added to substitute parent-ship (Red case). Tail position is assigned with skinning direction for the leaf bone (Blue case). 
    }
    \label{fig:bones}
    \vspace{-1.5em}
\end{figure}

%% file: supmat/fig/fig_animation.tex
\begin{figure*}[t]
    \centering
    \includegraphics[width=\linewidth]{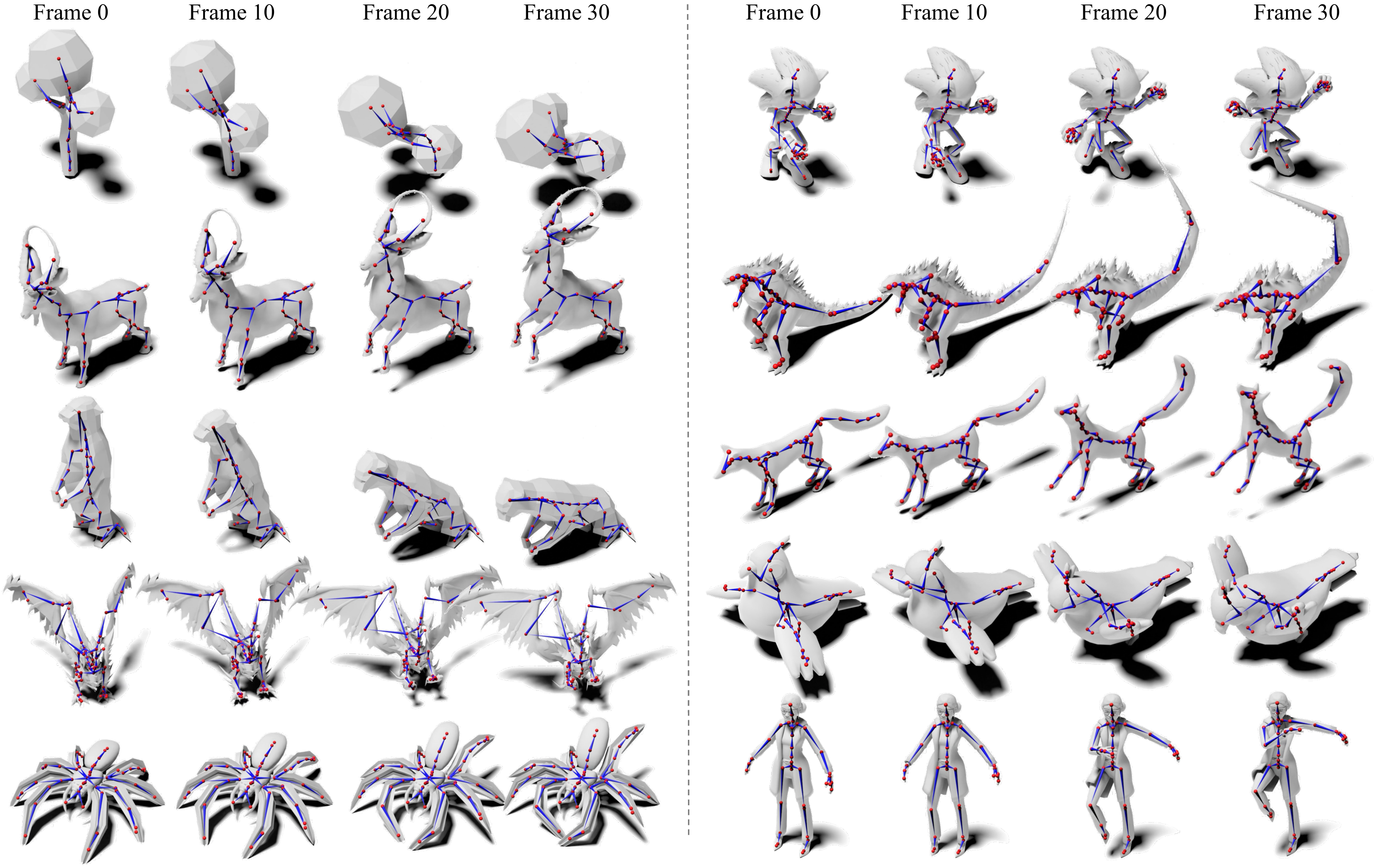}
    \vspace{-2.5em}
    \caption{Animation achieved using the rigging predicted by Anymate Model.
    }
    \label{fig:animation}
    \vspace{0 em}
\end{figure*}

%% file: supmat/fig/fig_diff.tex
\begin{figure}[t]
    \centering
    \includegraphics[trim={50pt 10pt 20pt 10pt}, clip, width=\linewidth]{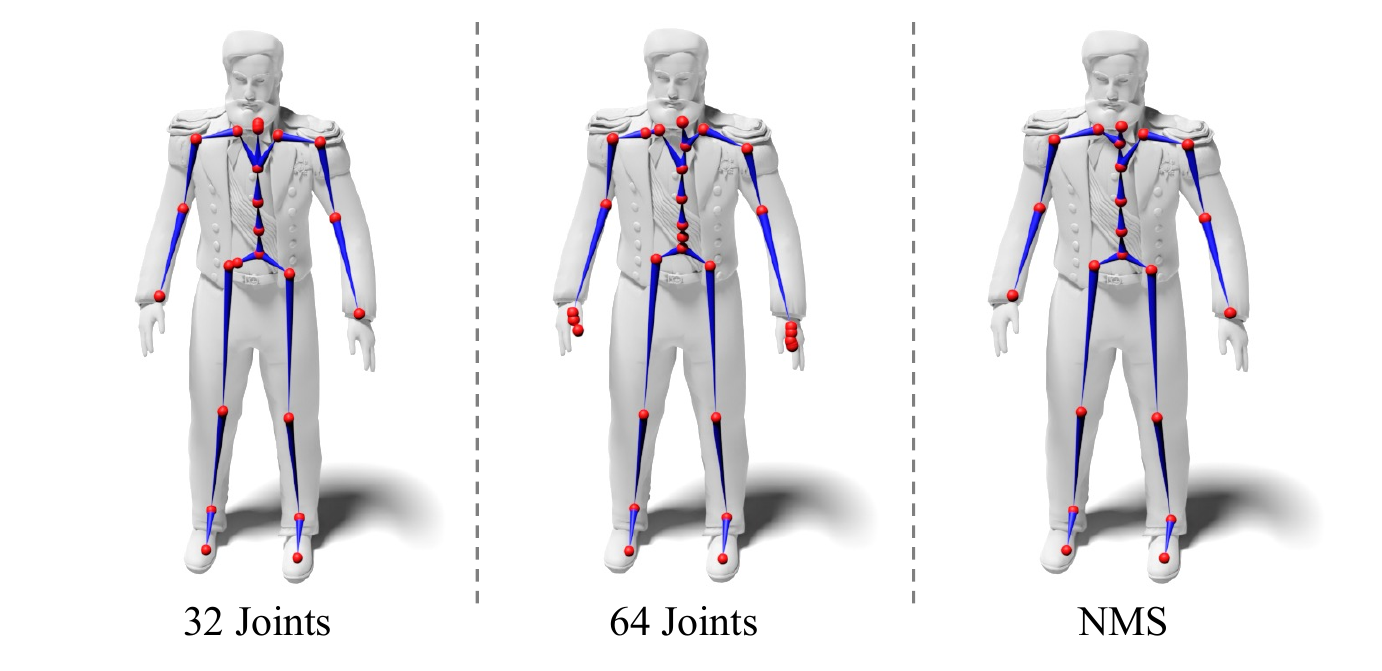}
    \caption{\textbf{Joint Prediction with Varying Numbers using the Diffusion-based Architecture.}
    Our proposed diffusion-based architecture allows user to specify a target number of joints, \eg 32 or 64, controlling different granularities of the predicted skeleton.
    We can also apply a non-maximum suppression (NMS) step which tends to retain a more cleaner set of joints, as shown on the right.
    }
    \label{fig:diff-vary}
\end{figure}

%% file: supmat/fig/fig_generated.tex
\begin{figure}[t]
    \centering
    \includegraphics[width=\linewidth]{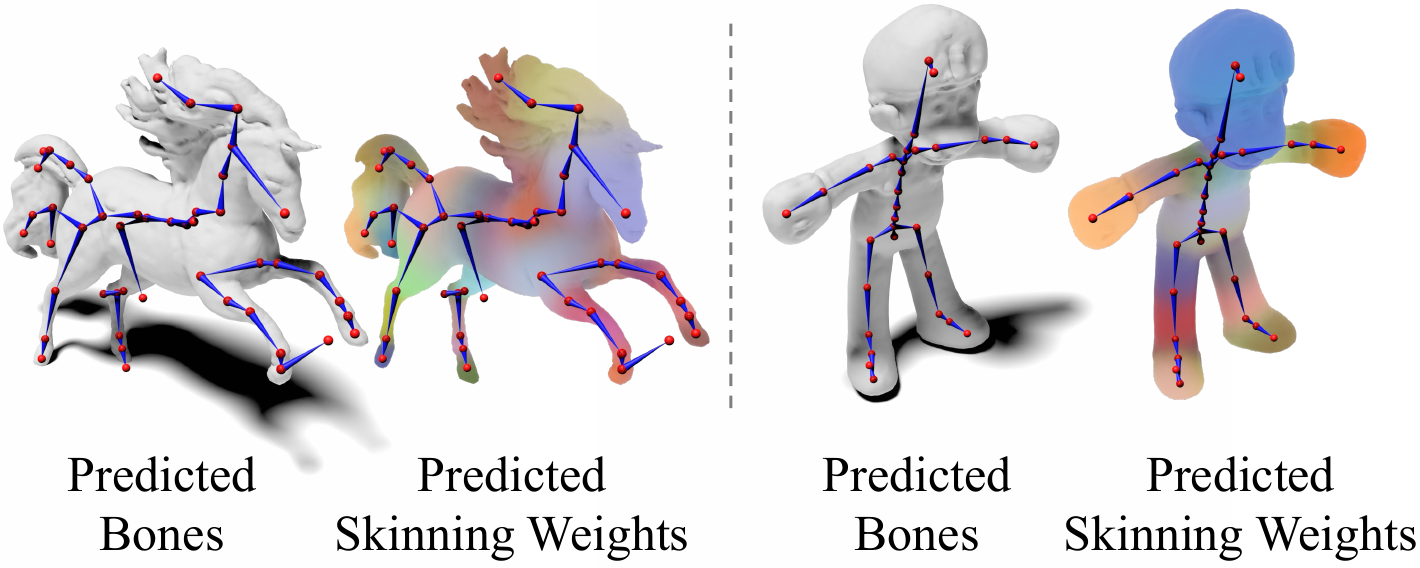}
    \caption{\textbf{Prediction Results on Meshes Generated by TripoSR~\cite{tochilkin2024triposr}}, which demonstrates the generalization capability of our model.
    }
    \label{fig:generated}
\end{figure}

%% file: supmat/tab/result_joint_supp.tex
\begin{table}[t]
    \small
    \centering
    \resizebox{\linewidth}{!}{%
    \begin{tabular}{lcccc}
        \toprule
        Model  & CD $\downarrow$ & EMD $\downarrow$ & Precision $\uparrow$ & Recall $\uparrow$\\
        \midrule
        Pinocchio~\cite{baran2007automatic}                     &0.198 & 0.659&52.7\% &24.9\% \\
        RigNet$^\dagger$~\cite{xu2020rignet}                           &0.144 &0.173 &54.6\% &57.1\% \\
        \midrule
        \multicolumn{4}{l}{\ \ \textit{trained on \datasetacronym-small (14K)}} \\
        RigNet~\cite{xu2020rignet}                       & 0.139& 0.166 & 49.4\% & 51.6\%\\
        Mich-Regress                  &0.099& 0.124& 71.6\%&70.9\% \\
        Mich-Diff$_\mathrm{Concat}$ & 0.142& 0.175&42.2\% & 50.9\% \\
        Mich-Diff$_\mathrm{Cross}$ & 0.128& 0.160& 65.1\%&64.9\% \\
        Mich-Vol$_\mathrm{Implicit}$  & 0.147& 0.127&70.3\% &73.7\% \\
        Mich-Vol$_\mathrm{TriPlane}$  & 0.198& 0.186&56.6\% & 49.2\%\\
        Bert-Regress                  & \textbf{0.091}& \textbf{0.113}& \textbf{74.7}\% & \textbf{79.8}\% \\
        Bert-Diff$_\mathrm{Concat}$ & 0.147& 0.177& 43.9\% & 58.3\%\\
        Bert-Diff$_\mathrm{Cross}$& 0.137& 0.167& 36.9\%& 30.5\% \\
        Bert-Vol$_\mathrm{Implicit}$  & 0.123& 0.126& 73.2\%& 69.9\% \\
        Bert-Vol$_\mathrm{TriPlane}$  & 0.192 &0.171& 63.8\% & 50.9\%\\
        \midrule
        \multicolumn{3}{l}{\ \ \textit{trained on \datasetacronym-train (225K)}} \\
        RigNet~\cite{xu2020rignet}                       &0.089 &0.127& 53.4\% &51.2\%  \\
        Bert-Regress                  & \textbf{0.077}& \textbf{0.098}& \textbf{80.2}\%&\textbf{84.4}\% \\
        Mich-Diff$_\mathrm{Cross}$& 0.085& 0.107& 76.9\% & 81.3\% \\
        \bottomrule
    \end{tabular}
    }
    \caption{\textbf{Quantitative Evaluation of Joint Prediction on \dataset-test} corresponding to Tab.~2 of the main paper.
    $^\dagger$Model weights released by RigNet~\cite{xu2020rignet} trained on ModelsResource-RigNetv1~\cite{xu2019predicting}.
    }
    \label{tab:supp_joint}
\end{table}

%% file: supmat/tab/result_skin_supp.tex
\begin{table}[t]
    \small
    \centering
    \resizebox{\linewidth}{!}{%
    \begin{tabular}{lccccc}
        \toprule
        Model    & CE $\downarrow$ & Cos $\uparrow$ & MAE $\downarrow$ & Precision $\uparrow$ & Recall $\uparrow$  \\
        \midrule
        GeoVoxel~\cite{dionne2013geodesic}                    & 3.569& 0.407& 0.056& 27.8\%& 34.6\%\\
        RigNet$^\dagger$~\cite{xu2020rignet}                               & 1.672 &0.661 & 0.029 &37.2\% &89.9\% \\
        \midrule
        \multicolumn{4}{l}{\ \ \textit{trained on \datasetacronym-small (14K)}} \\
        RigNet~\cite{xu2020rignet}                      & 1.573& 0.671&0.031 &37.2\% & 89.9\%   \\
        Mich-Concat                    & 1.169& 0.754& 0.029 &59.4\%  &87.1\% \\
        Mich-Cross                     & 1.270&0.766 & 0.026 &67.9\%  &86.8\%\\
        Bert-Concat                      & 1.242& 0.768&0.027&61.2\% & \textbf{90.2}\%  \\
        Bert-Cross                       & \textbf{1.134}& \textbf{0.801}& \textbf{0.025}& \textbf{68.8}\% &88.5\% \\
        \midrule
        \multicolumn{4}{l}{\ \ \textit{trained on \datasetacronym-train (225K)}} \\
        RigNet~\cite{xu2020rignet}                   & 1.521& 0.693& 0.028&37.2\% & 89.9\%    \\
        Bert-Cross                & \textbf{0.741} & \textbf{0.915}& \textbf{0.015} &\textbf{82.4\%}  & \textbf{93.2\%}\\ 
        \bottomrule
    \end{tabular}
    }
    \caption{\textbf{Quantitative Evaluation of Skinning Weight Prediction on \dataset-test} corresponding to Tab.~4 of the main paper.
    $^\dagger$Model weights released by RigNet~\cite{xu2020rignet} trained on ModelsResource-RigNetv1~\cite{xu2019predicting}.
    }
    \label{tab:supp_skin}
\end{table}

%% file: supmat/tab/result_joint_RigNetTest.tex
\begin{table}[t]
    \small
    \centering
    \resizebox{\linewidth}{!}{%
    \begin{tabular}{lcccc}
        \toprule
        Model  & CD $\downarrow$ & EMD $\downarrow$ & Precision $\uparrow$ & Recall $\uparrow$\\
        \midrule
        Pinocchio~\cite{baran2007automatic}                     & 0.213&0.512 & 36.6\% &25.1\% \\
        RigNet$^\dagger$~\cite{xu2020rignet}           &0.092 &0.124 &67.4\% & 75.7\%\\
        \midrule
        \multicolumn{5}{l}{\ \ \textit{trained on \datasetacronym-small (14K)}} \\
        RigNet~\cite{xu2020rignet}                       &0.108 &0.179  &63.8\% &48.5\% \\
        Mich-Regress                  &0.081& 0.099& \textbf{76.6}\%&73.5\% \\
        Mich-Diff$_\mathrm{Concat}$ & 0.117& 0.137& 55.4\%& 61.2\% \\
        Mich-Diff$_\mathrm{Cross}$ & 0.091& 0.101&  68.9\%& 69.7\%\\
        Mich-Vol$_\mathrm{Implicit}$  & 0.113& 0.121& 65.3\% &68.5\% \\
        Mich-Vol$_\mathrm{TriPlane}$  & 0.155& 0.167&49.2\% & 57.3\%\\
        Bert-Regress                  & \textbf{0.075}& \textbf{0.091}& 74.4\% & \textbf{80.4}\% \\
        Bert-Diff$_\mathrm{Concat}$ &0.134 &0.154 & 48.2\%  &53.9\% \\
        Bert-Diff$_\mathrm{Cross}$& 0.117& 0.137 & 53.4\% &65.6\%  \\
        Bert-Vol$_\mathrm{Implicit}$  & 0.105& 0.116& 69.8\%& 71.7\% \\
        Bert-Vol$_\mathrm{TriPlane}$  & 0.147 &0.162& 55.4\% & 59.0\%\\
        \midrule
        \multicolumn{5}{l}{\ \ \textit{trained on \datasetacronym-train (225K)}} \\
        RigNet~\cite{xu2020rignet}                       & 0.121& 0.291&53.6\%  &43.7\%  \\
        Bert-Regress                  & \textbf{0.057}& \textbf{0.068}& \textbf{85.6\%}& \textbf{88.9\%} \\
        Mich-Diff$_\mathrm{Cross}$& 0.070&0.088 & 78.7\% & 83.0\% \\
        \bottomrule
    \end{tabular}
    }
    \caption{\textbf{Quantitative Evaluation of Joint Prediction on RigNet-test~\cite{xu2020rignet}.}
    $^\dagger$Model weights released by RigNet~\cite{xu2020rignet} trained on ModelsResource-RigNetv1~\cite{xu2019predicting}.
    }
    \label{tab:supp_joint_rignet}
\end{table}

%% file: supmat/tab/result_con_RigNetTest.tex
\begin{table}[t]
    \small
    \centering
    \begin{tabular}{lcc}
        \toprule
        Model   & Precision $\uparrow$ & Recall $\uparrow$\\
        \midrule
        RigNet$^\dagger$~\cite{xu2020rignet}           &76.5\% &75.3\% \\
        \midrule
        \multicolumn{3}{l}{\ \ \textit{trained on \datasetacronym-small (14K)}} \\
        RigNet~\cite{xu2020rignet}        &73.2\% & 71.5\%\\
        Mich-Token        &65.3\% &63.5\% \\
        Mich-2Branch    & 73.7\%& 71.5 \% \\
        Bert-Token        &  69.9\%&66.8 \%\\
        Bert-2Branch    & \textbf{74.6\%}& \textbf{72.3\%}\\
        \midrule
        \multicolumn{3}{l}{\ \ \textit{trained on \datasetacronym-train (225K)}} \\
        RigNet~\cite{xu2020rignet}    &   68.8\% & 67.6\%\\
        Bert-2Branch    & \textbf{89.8\%}& \textbf{88.7\%}\\
        \bottomrule
    \end{tabular}
    \caption{\textbf{Quantitative Evaluation of Connectivity Prediction on RigNet-test~\cite{xu2020rignet}.}
    $^\dagger$Model weights released by RigNet~\cite{xu2020rignet} trained on ModelsResource-RigNetv1~\cite{xu2019predicting}.
    }
    \label{tab:supp_con_rignet}
\end{table}

%% file: supmat/tab/result_skin_RigNetTest.tex
\begin{table}[t]
    \small
    \centering
    \resizebox{\linewidth}{!}{%
    \begin{tabular}{lccccc}
        \toprule
        Model    & CE $\downarrow$ & Cos $\uparrow$ & MAE $\downarrow$ & Precision $\uparrow$ & Recall $\uparrow$  \\
        \midrule
        GeoVoxel~\cite{dionne2013geodesic}                    &3.012 &0.451 &0.071 &26.4\% &32.9\%\\
        RigNet$^\dagger$~\cite{xu2020rignet}     &1.054  &0.742 &0.056  & 34.9\% &93.7\% \\
        \midrule
        \multicolumn{6}{l}{\ \ \textit{trained on \datasetacronym-small (14K)}} \\
        RigNet~\cite{xu2020rignet}                      &1.122 &0.723 &0.059 &34.7\% &93.4\%    \\
        Mich-Concat                    & 0.758& 0.794& 0.034 &61.6\%  &93.1\% \\
        Mich-Cross                     &0.783&0.810 & 0.031 &68.2\%  &92.5\%\\
        Bert-Concat                      & 0.727& 0.821&0.032&63.9\% & \textbf{95.4\%}  \\
        Bert-Cross                       & \textbf{0.711}& \textbf{0.829}& \textbf{0.029}& \textbf{67.9}\% & 93.9\% \\
        \midrule
        \multicolumn{6}{l}{\ \ \textit{trained on \datasetacronym-train (225K)}} \\
        RigNet~\cite{xu2020rignet}                   &1.012 &0.744 &0.052& 34.8\% & 94.0\%   \\
        Bert-Cross                & \textbf{0.534} & \textbf{0.891}& \textbf{0.023} &\textbf{73.2\%}  & \textbf{96.0\%}\\ 
        \bottomrule
    \end{tabular}
    }
    \caption{\textbf{Quantitative Evaluation of Skinning Weight Prediction on RigNet-test~\cite{xu2020rignet}.}
    $^\dagger$Model weights released by RigNet~\cite{xu2020rignet} trained on ModelsResource-RigNetv1~\cite{xu2019predicting}.
    }
    \label{tab:supp_skin_rignet}
\end{table}

%% file: tab/Joints_RigNet.tex
\begin{table}[t]
    \small
    \centering
    \begin{tabular}{lcccc}
        \toprule
        Model  & CD $\downarrow$ & EMD $\downarrow$ & Precision $\uparrow$ & Recall $\uparrow$\\
        \midrule

        RigNet$^\dagger$~\cite{xu2020rignet}           &0.092 &0.124 &67.4\% & 63.7\%\\

        Mich-Regress$^{*}$                 & \textbf{0.023} & \textbf{0.034}& \textbf{90.7}\%& \textbf{82.1}\%\\
        Mich-Diff$_\mathrm{Cross}$$^{*}$& 0.032&0.041 & 81.7\% & 78.8\% \\
        \bottomrule
    \end{tabular}
    \caption{\textbf{Quantitative Evaluation of Joint Prediction on RigNet-test~\cite{xu2020rignet}.}
    $^\dagger$Model weights released by RigNet~\cite{xu2020rignet} trained on ModelsResource-RigNetv1~\cite{xu2019predicting}. $^{*}$Model trained on ModelsResource RigNetv1~\cite{xu2019predicting}.
    }
    \label{tab:rubb_joint_rignet}
    \vspace{-0.1in}
\end{table}

%% file: tab/Con_RigNet.tex
\begin{table}[t]
    \small
    \centering
    \resizebox{\linewidth}{!}{
    \begin{tabular}{lcccc}
        \toprule
        Model   &ED $\downarrow$& Precision $\uparrow$ & Recall $\uparrow$ & CD-B2B $\downarrow$ \\
        \midrule
        RigNet$^\dagger$~\cite{xu2020rignet}          &2.52 &84.5\% &77.3\% & 0.012\\
        Bert-2Branch$^{*}$    &\textbf{0.81} & \textbf{96.7}\%& \textbf{96.6}\% & \textbf{0.002}\\
        \bottomrule
    \end{tabular}
    }
    \caption{\textbf{Quantitative Evaluation of Connectivity Prediction on RigNet-test~\cite{xu2020rignet}.}
    $^\dagger$Model weights released by RigNet~\cite{xu2020rignet} trained on ModelsResource-RigNetv1~\cite{xu2019predicting}. $^{*}$Model trained on ModelsResource RigNetv1~\cite{xu2019predicting}.
    }

    \label{tab:rubb_con_rignet}
    \vspace{-0.1in}
\end{table}

%% file: tab/Skeleton_RigNet.tex
\begin{table}[t]
    \small
    \centering
    \begin{tabular}{lccc}
        \toprule
        Model  & IoU $\uparrow$ & CD-J2B $\downarrow$ & CD-B2B $\downarrow$ \\
        \midrule

        RigNet$^\dagger$~\cite{xu2020rignet}           &63.4\% &0.051 &0.050 \\

        Mich-Regress$^{*}$ + Bert-2Branch$^{*}$                  & \textbf{85.6}\% & \textbf{0.017}& \textbf{0.019}\\
        Mich-Diff$_\mathrm{Cross}$$^{*}$ + Bert-2Branch$^{*}$& 79.8\%&0.021 & 0.023 \\
        \bottomrule
    \end{tabular}
    \caption{\textbf{Quantitative Evaluation of Skeleton Prediction on RigNet-test~\cite{xu2020rignet}.}
    $^\dagger$Model weights released by RigNet~\cite{xu2020rignet} trained on ModelsResource-RigNetv1~\cite{xu2019predicting}. $^{*}$Model trained on ModelsResource RigNetv1~\cite{xu2019predicting}.
    }
    \label{tab:rubb_skel_rignet}
    \vspace{-0.1in}
\end{table}

%% file: tab/Skin_RigNet.tex
\begin{table}[t]
    \small
    \centering
    \resizebox{\linewidth}{!}{%
    \begin{tabular}{lccccc}
        \toprule
        Model    & CE $\downarrow$ & Cos $\uparrow$ & L1-avg $\downarrow$ & Precision $\uparrow$ & Recall $\uparrow$  \\
        \midrule
        RigNet$^\dagger$~\cite{xu2020rignet}     &1.054  &0.742 &0.57  & 60.9\% &89.7\% \\

        Bert-Cross$^{*}$               & \textbf{0.458} & \textbf{0.922}& \textbf{0.26} &\textbf{79.4}\%  & \textbf{96.4}\%\\ 
        \bottomrule
    \end{tabular}
    }
    \caption{\textbf{Quantitative Evaluation of Skinning Weight Prediction on RigNet-test~\cite{xu2020rignet}.}
    $^\dagger$Model weights released by RigNet~\cite{xu2020rignet} trained on ModelsResource-RigNetv1~\cite{xu2019predicting}. $^{*}$Model trained on ModelsResource RigNetv1~\cite{xu2019predicting}.
    }
    \label{tab:rubb_skin_rignet}
    \vspace{-0.1in}
\end{table}

%% file: tab/implementation.tex
\begin{table}[t]
    \small
    \centering
    \begin{tabular}{lc}
        \toprule
        Training Hyper-Parameter  & Value \\
        \midrule
        Epoch & 200 \\
        Batch Size & 16 \\
        Learning Rate         & 1e-4\\
        Learning Rate Decay            & 2e-5 at epoch 100\\
        Optimizer & Adam\\
        \bottomrule
    \end{tabular}

    \vspace{1em}

    \begin{tabular}{lcc}
        \toprule
        Model  & Layer Number \& Type & Layer Dimensions \\
        \midrule
        Encoder-Mich            & 25 TFLayer & 768\\
        Encoder-Bert            & 12 TFLayer & 384\\
        Joint-Regress & 1 TFLayer & 768 \\
        Joint-Diff$_\mathrm{Concat}$ & 4 TFLayer & 512\\
        Joint-Diff$_\mathrm{Cross}$ & 4 DenoiseBlock & 768\\
        Joint-Vol$_\mathrm{Implicit}$  & 1 TFLayer & 768\\
        Joint-Vol$_\mathrm{TriPlane}$  & 7 CNNBlock & 512\\
        Conn-Token & 4 TFLayer & 768\\
        Conn-2Branch & 5 TFLayer & 768\\
        Skin-Concat  & 4 TFLayer &768\\
        Skin-Cross  & 4 TFLayer & 768 \\
        \bottomrule
    \end{tabular}
    \caption{\textbf{Implementation Details:} Hyper-Parameter and Model Size. In the model details table, "TF" denotes "Transformer", and "Layer Dimensions" represents the size of the token used by Transformer or the majority channel size used by CNNBlock.
    }
    
    \label{tab:implementation}
    \label{tab:impl}
\end{table}

%% file: supmat/fig/fig_failure.tex
\begin{figure}[t]
    \centering
    \includegraphics[width=\linewidth]{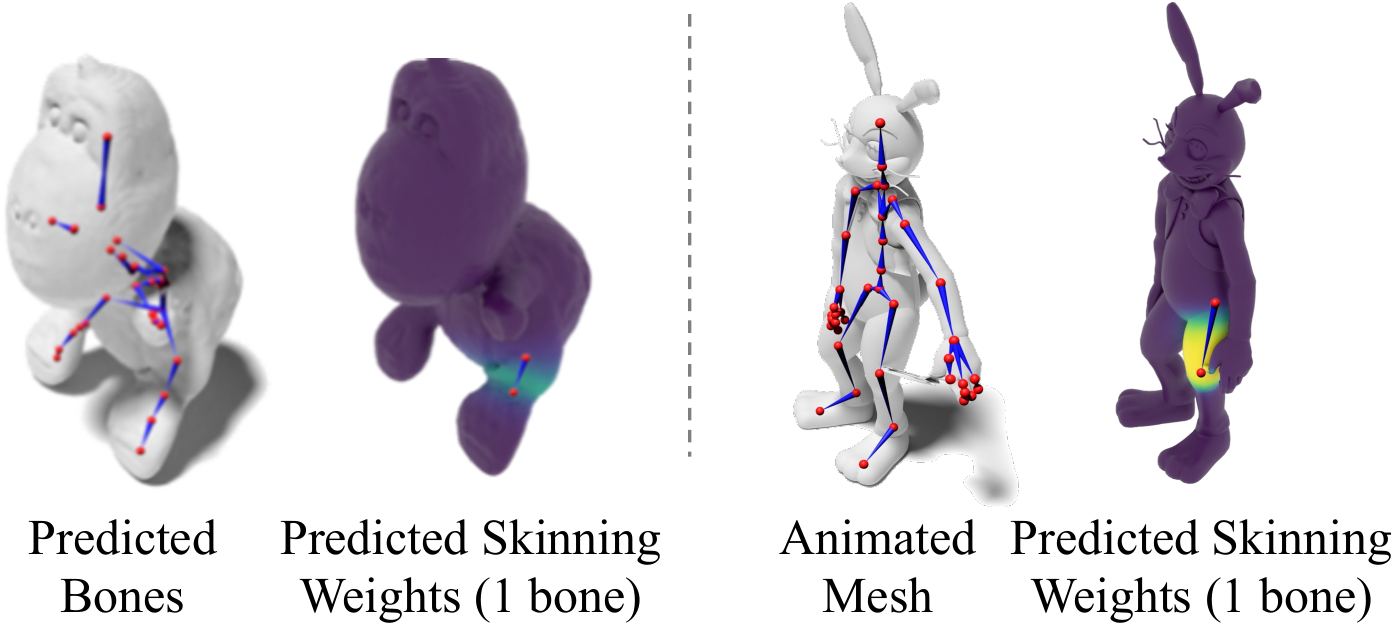}
    \caption{\textbf{Failure Cases.}
    On the left, we show a failure case where the predicted bones do not form a reasonable skeleton.
    On the right, we show another example where the predicted skinning weights incorrectly associate the finger to thigh bone, resulting in distortion when animated.
    }
    \label{fig:failure}
\end{figure}